\documentclass[preprint]{aastex}
\usepackage{amsmath}

\shorttitle{Polytropic Stars in Close Binaries}
\shortauthors{Sirotkin \& Kim}

\newcommand {\I} {{\small{I}}}
\newcommand {\II} {{\small{II}}}
\newcommand {\Jspin} {J_{\rm spin}}
\newcommand {\Jorb} {J_{\rm orb}}
\newcommand {\vel} {V}
\newcommand {\trho} {\Theta^n}
\newcommand {\brho} {\bar\rho}
\newcommand {\apside} {\dot\varpi}
\newcommand {\mumin} {\mu_{\rm min}}
\newcommand {\keff} {k_{2,\rm eff}}
\newcommand {\kiso} {k_{2, \rm iso}}
\newcommand\simgt{\lower.5ex\hbox{$\; \buildrel > \over \sim \;$}}
\newcommand\simlt{\lower.5ex\hbox{$\; \buildrel < \over \sim \;$}}

\begin{document}
\title{Internal Structure and Apsidal Motions of Polytropic Stars in Close 
Binaries}
\author{Fedir V.\ Sirotkin \& Woong-Tae Kim}
\email{sirotkin.f.v@gmail.com, wkim@astro.snu.ac.kr}
\affil{Department of Physics and Astronomy, FPRD, 
Seoul National University, Seoul, 151-742, South Korea} 

\begin{abstract}
We consider a synchronized, circular-orbit binary consisting of 
a polytrope with index $n$ and a point-mass object, and use a self-consistent 
field method to construct the equilibrium structure of the polytrope 
under rotational and tidal perturbations.
Our self-consistent field method is distinct from others in that the 
equilibrium orbital angular velocity is calculated automatically rather than 
being prescribed, which is crucial for obtaining apsidal motion rates 
accurately.  We find that the centrifugal and tidal forces make perturbed 
stars more centrally condensed and larger in size.  For
$n=1.5$ polytopes with fixed entropy,  the enhancement factor in stellar
radii is about $23\%$ and $4-8\%$ for $\mu=1$ and $\sim0.1-0.9$, respectively, 
where $\mu$ is the fractional mass of the polytrope relative to the total. 
The centrifugal force dominates the tidal force in determining the
equilibrium structure provided $\mu \simgt 0.13-0.14$ for $n\simgt 1.5$.
The shape and size of rotationally- and tidally-perturbed polytropes 
are well described by the corresponding Roche models as long as $n\simgt 2$.
The apsidal motion rates calculated for circular-orbit binaries 
under the equilibrium tide condition agree well with the predictions 
of the classical formula only when the rotational and tidal perturbations 
are weak.  When the perturbations are strong as in critical configurations,
the classical theory underestimates the real apsidal motion rates by
as much as 50\% for $n=1.5$ polytropes, although the discrepancy 
becomes smaller as $n$ increases.  For practical uses, we provide fitting 
formulae for the density concentration, volume radius, coefficient
of the mass-radius relation, moment of inertia, spin angular momentum, 
critical rotation parameter, effective internal structure constant, etc.,
as functions of $\mu$ and the perturbation parameters.
\end{abstract}


\section{Introduction} \label{sec_introduction}

Binary systems are an ideal laboratory for the study of stellar 
astrophysics.
They not only provide direct information on stellar mass, radius, and 
luminosity,  which are all fundamental for the theory of stellar 
evolution, but also allow to probe stellar structures by measuring
apsidal motions (e.g., \citealt{egg06,gim07}).
Since tidal forces from a star in a binary system are likely to 
affect the rotation rate and internal structure of the other component, 
one expects that the stellar quantities such as radius and
effective temperature of binary stars may differ from those of single stars,
as confirmed by observations (e.g., \citealt{mal03,mal07,rib06}).
While the literature abounds with studies on the rotational and tidal 
distortions of stars, most of the studies are either approximate or based on 
prescribed rotation profiles, as detailed below. 
In this work, we employ a fully self-consistent method to investigate 
the quantitative effects of centrifugal and
tidal forces on equilibrium polytropes in synchronously-rotating, 
circular-orbit binaries. 

Finding equilibrium configurations of polytropic stars under
rotational (and/or tidal) distortion dates back to
\citet{cha33a,cha33b,cha33c} who used a perturbation theory in which
an equilibrium density distribution was regarded as a small departure 
from a corresponding unperturbed Lane-Emden function.  
Chandrasekhar considered only first-order terms in the rotation parameter 
$\upsilon$ (see eq.\ [\ref{eq_ups_mu_nu}] for definition),  
corresponding to slow rotation. Later, this perturbation method 
was extended to include terms of second order \citep{ana68, ger79} and  
third order \citep{ger85}.  
The first-order perturbation theory was further refined by  
\citet{lin77,lin81} who retained the first-order terms 
in the potential expansion, while allowing for large deformation of
the equilibrium configurations from the unperturbed polytropes.  
More recently, \citet{ger88} showed that the 
first-order perturbation theory combined with ``complex Lane-Emden 
functions'' beyond their first roots 
gives much improved results for rotating polytropes.

\citet{kip70} used a quasi-spherical method in which (1) the rotation
velocity is assumed to be spherically symmetric rather than 
being cylindrically symmetric and (2) only the radial 
component of the centrifugal force is considered in the force balance;
these assumptions make rotating stars essentially spherical.
Recently, \citet{arb09} used a similar method to study the change of the total 
angular momentum due to tidal perturbations and the associated tidal
instability.
While the quasi-spherical method was able to handle the effects of rotation
on stellar evolution in a very simplified way, it readily fails when the 
stellar shape becomes quite non-spherical.

To study the internal structure of binary stars,  
\citet{kop72} constructed the Roche model where the actual 
equipotential surfaces of distorted stars are approximated by the 
corresponding Roche potentials of two point-mass stars
(see also \citealt{kop78,moh78,moh83,moh90,moh92,moh97}).
While the Roche model is a good approximation for stars 
with very high central
concentration, it assumes Keplerian orbits and thus cannot give any clue to 
the apsidal motions of binary stars.

On the other hand, \citet{rox65} developed a so-called 
``double-approximation method'' in which a distorted star is divided
into two parts: a core containing most of the stellar material and 
a non-self-gravitating, low-density envelope.
While the envelope is highly susceptible to applied centrifugal 
and tidal forces, the distortion of the core is assumed to be small
even in the case of critical rotation. 
Depending on the method to calculate the distortion of the core,
the double-approximation method has several different branches:
\citet{rox65}, \citet{jac70}, \citet{dur70}, and \citet{nay72a,nay72b} 
used first-order perturbation theory of \citet{cha33a}; 
\citet{mar70} and \citet{sin84} included terms up to of second order; 
\citet{may97} and \citet{lal06} employed the quasi-spherical method.

Although the aforementioned work have improved our understanding of 
the rotational and tidal effects on stellar structure, these 
are only approximate and do not yield accurate 
results when the perturbations are rather strong.
It was \citet{jam64} who first sought for direct numerical solutions 
for rotationally distorted polytropes without any assumption. 
By expanding the potential-density pairs in 
Legendre polynomials up to of tenth order and integrating a set of
the resulting ordinary differential equations \emph{simultaneously},
\citet{jam64} obtained solutions, with truncation errors less 
than 0.2\%, for rotating polytropes.  On the other hand,
\citet{ost68} introduced a self-consistent field 
(SCF) method that provides a concise, accurate, and very efficient 
computation scheme. 
In the SCF method, the equilibrium density distribution is found 
by solving the Poisson equation as well as the equilibrium condition, 
\emph{alternatively} and \emph{iteratively}, for the potential-density pairs. 
The SCF method is so powerful that it has been applied to various
problems including the structure of rotating stars 
\citep{ost_bod68,mar68,cle74,sta83a,sta83b,hac86a} and 
galactic dynamics of collisionless systems \citep{hern92,her95,wei99,gne03}.
Recently, \citet{mac07} used a modified version of the
SCF method to construct stellar models for 
differentially-rotating, low-mass, main-sequence stars with a realistic 
equation of state  (see also \citealt{jac05}).

While most SCF methods for distorted stars are two-dimensional
under rotational symmetry, 
\citet{hac86b} formulated a three-dimensional SCF method
to calculate structures of rotating multi-body systems in 
which tidal effects are implicitly included.
He showed numerically that there is an equilibrium sequence 
along which an ellipsoid takes on a dumbbell-like shape as the angular 
momentum of the whole system increases,  eventually separating into
a binary.
Although \citet{hac86b} calculated the density structures and shapes of 
multi-body systems, these are for equal-mass objects (owing to the imposed
azimuthal symmetry) with a few prescribed laws for spin and orbital 
angular velocities. 
Later, \citet{hacetal86a,hacetal86b} generalized the method 
to allow for unequal-mass binaries, but the rotational velocity is still 
given externally rather than being calculated self-consistently,
which makes it impossible to obtain information on apsidal motions.

Finding accurate internal structures of distorted stars in close binary 
systems are also of great importance in testing the classical theory of 
apsidal motions.  Apsidal motions arise when the gravitational force due
to a star is different from that of a Newtonian point-mass.  The chief cause
of perturbations to Keplerian orbits is the rotational and tidal distortions
that make the figures of binary stars deviate from spherical symmetry.
Also, the separate or combined effects of the 
general relativistic correction, the presence of a third body, and 
non-equilibrium tides can be also non-negligible
(see recent review of \citealt{gim07}).  
For binary stars in near-circular motions, 
\citet{rus28} showed that the 
advance rate, $\apside$, of the apsidal line due to the tidal and 
rotational distortion of star \I\ with mass $M_1$ 
caused by the attraction of star \II\ with mass $M_2$ is given by 
\begin{equation}\label{eq_apsidal}
\frac{\apside}{\Omega_K} = k_2
\Biggl(\frac{R_1}{A}\Biggr)^{5}
\Biggl({1+7\frac{M_2} {M_1}} \Biggr),
\end{equation}
where $\Omega_K=(G[M_1+M_2]/A^3)^{1/2}$ is the Keplerian angular velocity, 
$R_1$ is the mean radius of star \I,
$A$ is the orbital separation, and $k_2$ is the 
dimensionless apsidal motion constant. 
Through more rigorous, first-order perturbation analyses, 
\citet{cow38} and \citet{ste39} generalized
equation (\ref{eq_apsidal}) to make it applicable to binaries with
eccentric orbits, which has been widely used to estimate 
the degree of the internal density concentration via $k_2$ from observations. 

However, the classical formulae of \citet{rus28}, \citet{cow38}, and 
\citet{ste39} have two limitations.  First, by ignoring the cross 
terms of spherical harmonics in expanding equipotential surfaces, 
they are valid in a strict sense only when 
the tidal and rotational distortions are weak.  Second, by 
assuming an instantaneous adjustment of a star to time-dependent tidal force,
they fail when the orbital period is shorter than the oscillation periods of
tidally excited modes inside the perturbed star.  \citet{qua96} and
\citet{sme01} relaxed the second limitation to study the effects of dynamical 
and resonance tides (still based on a linear perturbation theory), 
finding that the classical formulae are accurate as 
long as the ratio of the modal to orbital frequencies is less than about 10
(see also \citealt{cla03}).  
The validity of the classical formulae is nonetheless questionable 
for close binaries, especially in the critical configurations,  
where the distortion of a stellar shape is large enough for
nonlinear terms to be non-negligible.

Despite a long history of research, therefore, studies are still lacking 
as to how the stellar structure, shape, size, etc.\ of a distorted star
as well as its apsidal motion rate in a close binary system vary with 
the fundamental parameters such as the mass ratio and orbital distance. 
More specifically, 
how good are the approximate methods described above for highly
distorted stars?
What is the relative role of tidal force to centrifugal force in 
increasing the central density concentration?
Can the classical theory of apsidal motions based on perturbation 
analyses be accurately applied even to critical configurations? 
To address these questions, we in this work consider a 
circular-orbit binary in which two components are modeled as 
a polytrope and a point-mass, respectively,
and calculate the equilibrium structure of the 
polytrope using an SCF method.  The binary is assumed to be in
spin-orbit synchronization, so that the tide is independent of time. 
As explained below,
two salient features of our SCF method compared to those in the previous 
studies are that the former considers only the polytropic component 
in density computation, while explicitly including the tidal force from
its companion as a fixed gravitational potential,
and that the orbital velocity 
indispensable for calculating the apsidal motion rates
is now obtained self-consistently from the equilibrium condition.

The organization of this paper is as follows.
In \S\ref{sec_for}, we describe our implementation of the SCF method to 
find the equilibrium structure of a synchronously rotating binary 
component.  In \S3, we first present the numerical solutions for unperturbed
polytropes and uniformly-rotating polytropes and compare them with the 
published results.  We then turn to the solutions for both 
rotationally- and tidally-distorted polytropes, and give detailed 
quantitative results on the changes of internal structures 
as well as on the parameters describing critical configurations.
Using the self-consistent solutions of perturbed polytropes, we test
the classical theory of apsidal motions in \S4. 
Finally in \S5, we summarize our results
and discuss the astronomical implication of our work.

\section{Formulation}\label{sec_for}

\subsection{Basic Equations} \label{sec_stell_struct}

\begin{figure*}[c]
\epsscale{1.}
\plotone{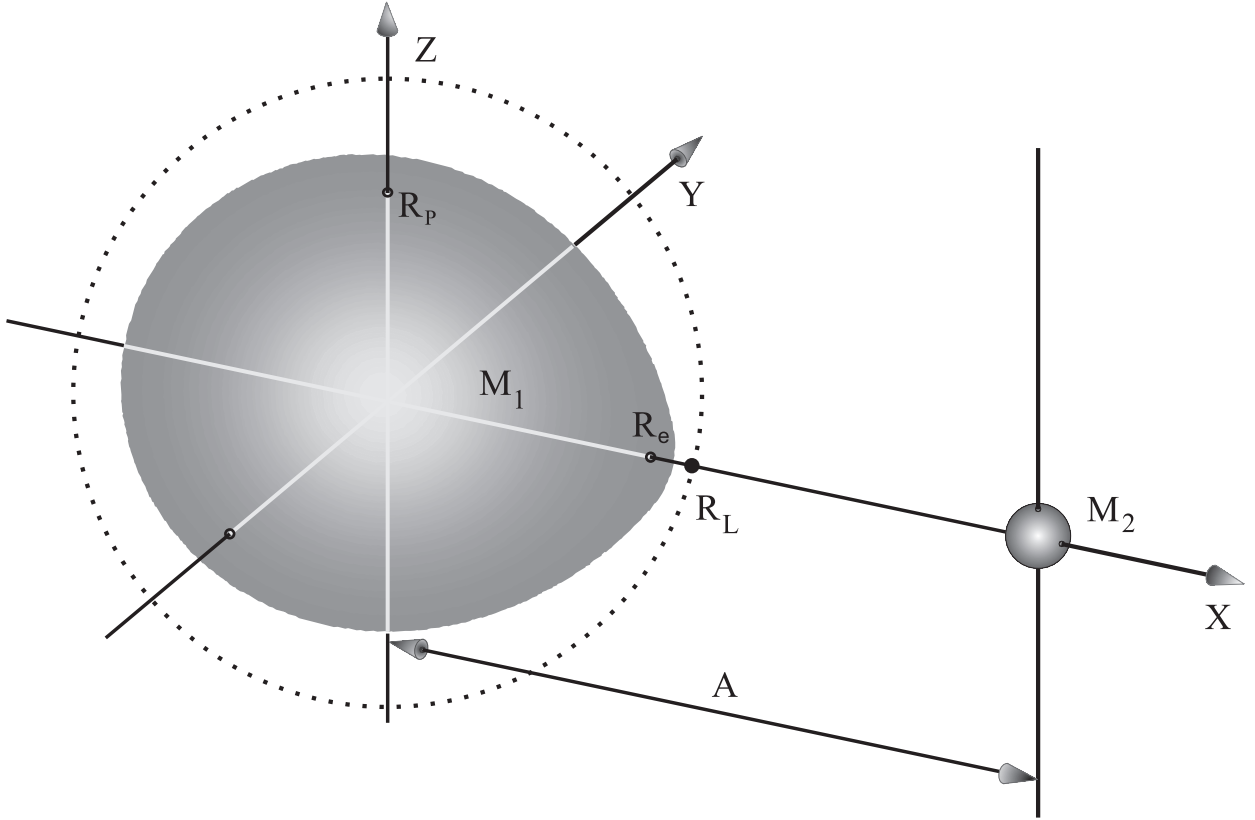}
\caption{
Schematic view of a binary configuration 
consisting of a distorted polytrope (star \I) with mass $M_1$ 
and its point-mass companion (star \II) with mass $M_2$.  
The Cartesian coordinate system $(x, y, z)$ corotating with 
the binary is centered  
at the center of mass of star \I\ that has polar and equatorial 
radii $R_p$ and $R_e$ in the $z$- and positive $x$-directions, 
respectively.  Star \II\ is located at $(A, 0, 0)$.  Dotted line draws 
a sphere with radius equal to the distance from the center of mass of
star \I\ to the Lagrange point $R_L$ of the binary. 
} \label{pic_geom}
\end{figure*}

In this paper we use an SCF method to study the equilibrium 
structure of a self-gravitating star under the external 
gravitational force from a companion in a close binary system.  
We assume that two stars revolve in circular orbits with
constant angular speed $\Omega$ and orbital separation $A$, so that 
the tidal force is stationary. 
We further assume that a star in question with
mass $M_1$ and volume radius $R_1$ (hereafter star \I),
composed of inviscid polytropic gas without magnetic fields,
rotates rigidly and synchronously with the orbital motion. 
For simplicity, we treat the companion star 
(hereafter star \II) as a point mass with mass $M_2$.
We work in a rotating Cartesian frame at $\Omega$ whose
origin lies at the center of mass of star \I.
Figure \ref{pic_geom} schematically illustrates the binary
configuration and the coordinate system in which the 
$x$-axis points toward star \II\ located at 
$(x,y,z)=(A,0,0)$ and the $z$-axis is perpendicular  to
the  orbital  plane.  
Star \I\ is distorted rotationally and tidally, and has a polar 
radius $R_p$ and an equatorial radius $R_e$ along the positive $x$-axis.
We assume for 
computational convenience that all the substance of star \I\ is contained 
within a sphere with radius $R_L$, the distance to the inner Lagrange point
from the origin.  

The equation of hydrostatic equilibrium then takes the form
\begin{equation}\label{eq_intro_equilibrium_cond}
\frac{\nabla P} {\rho}+\nabla\Phi=0,
\end{equation}
where $P$ is the pressure,  $\rho$ is the density, and
$\Phi$ is the effective potential defined by
\begin{equation}\label{eq_pot}
\Phi = \Phi_{\Omega} + \Phi_* + \Phi_g.
\end{equation}
In equation (\ref{eq_pot}), 
$\Phi_{\Omega}$ is the centrifugal potential
\begin{equation}\label{eq_rot}
\Phi_{\Omega}=-\frac{\Omega^{2}} {2} \{[x-A(1-\mu)] ^{2}+y^{2}\},
\end{equation}
with $\mu= M_1/(M_1 + M_2)$ being the mass fraction of star \I,
$\Phi_{*}$ is the gravitational potential from star \II\
\begin{equation}\label{eq_ext}
\Phi_{*}=-\frac{GM_{2}} {\sqrt{(x-A) ^{2}+y^{2}+z^{2}}},
\end{equation}
and $\Phi_g$ is the 
self-gravitational potential of star \I\ that 
satisfies the Poisson equation
\begin{equation} \label{eq_Poisson}
\nabla^{2} \Phi_{g}=4\pi G \rho.
\end{equation}
The proper boundary conditions on $\Phi_g$ are 
that $\Phi_{g}$ and its first derivatives should behave well
at $r=(x^2+y^2+z^2)^{1/2} \rightarrow 0$  and $r\rightarrow \infty$,
and should be continuous on the surface of a polytrope.  These
boundary conditions are automatically satisfied when we use 
the integral representation of the Poisson equation in
\S\ref{sec:exp} below.

We assume that the matter of star \I\ obeys a polytropic relation
\begin{equation} \label{eq_state}
 P=K \rho^{1+1/n},
\end{equation}
with the pressure constant $K$ and the polytropic index $n$.
Equation (\ref{eq_intro_equilibrium_cond}) is then integrated to yield
\begin{equation} \label{eq_balance}
\Phi+(n+1) K \rho^{1/n}=\Phi_s,
\end{equation}
where  $\Phi_s$ is the potential on the surface of star \I\
with vanishing density.
An equilibrium condition for star \II\ is simply 
\begin{equation} \label{eq_ang_vel}
\Omega^2 = \frac{1}{\mu A} 
\frac{\partial \Phi_g} {\partial x}
\Biggr|_{\mathbf{x}=(A, 0, 0)},
\end{equation}
which must be checked a posteriori.

Following the convention, we introduce the dimensionless
coordinates $\boldsymbol\xi=(\xi, \eta, \zeta)\equiv (x,y,z)/\alpha$, where 
\begin{equation}\label{eq_alpha}
  \alpha=\Biggl[\frac{(n+1)K} {4\pi G} \rho_c^{1/n-1}  \Biggr]^{1/2}.
\end{equation}
We also define the generalized Lane-Emden function $\Theta$ through
\begin{equation} \label{eq_den_dim}
  \rho=\rho_c\Theta^n,
\end{equation}
where the central density $\rho_c$ is given by
\begin{equation} \label{eq_cden}
\rho_c = \frac{M_1} {\alpha^3\int\Theta^{n} d^3\xi}.
\end{equation}
The equilibrium condition (\ref{eq_balance}) then becomes
\begin{equation} \label{eq_balance_dim}
\Theta=\Psi_s-\Psi,
\end{equation}
where $\Psi \equiv \Phi /([n+1]K\rho_c^{1/n})$ 
and $\Psi_s$  is the value of $\Psi$ on the surface of star \I.

In dimensionless forms, equations (\ref{eq_rot})-(\ref{eq_Poisson})
and (\ref{eq_ang_vel}) read
\begin{equation}\label{eq_rot_dless}
\Psi_{\Omega}=-\frac{\upsilon} {4} 
\{[{\xi}-\nu(1-\mu)]^2+\eta^2\} 
\end{equation}
\begin{equation}\label{eq_ext_dless}
\Psi_{*}=-\frac{\upsilon} {2} 
\Biggl(\frac{\Omega_{K}} {\Omega_{~}} \Biggr)^2
\frac{\nu^3(1-\mu)}{\sqrt{(\xi-\nu)^2+\eta^2+\zeta^2}},
\end{equation}
and
\begin{equation}\label{eq_Pos_dless}
\tilde{\nabla}^2 \Psi_g = \Theta^n,
\end{equation}
where $\tilde{\nabla}$ denotes a dimensionless derivative with
respect to $\boldsymbol\xi$. 
In equations (\ref{eq_rot_dless}) and (\ref{eq_ext_dless}),
$\Omega_K^2=G(M_1+M_2)/A^3$ is the Keplerian angular velocity,
$\nu\equiv A/\alpha$ is the dimensionless orbital distance,
and  $\upsilon$  is the dimensionless angular velocity defined by
\begin{equation}\label{eq_ups_mu_nu}
\upsilon\equiv \frac{\Omega^2} {2\pi G \rho_c}
=\Biggl(\frac{\Omega} {\Omega_K} \Biggr)^2\frac{\int\Theta^n d^3\xi}
{2\pi\mu\nu^3}
\end{equation}
which is a measure of the amplitude of rotational perturbations
(e.g., \citealt{cha33b}).  
It is sometimes useful to measure the degree of rotational perturbations 
in terms of the mean density $\brho\equiv 3M_1/(4\pi R_1^3)$ as 
\begin{equation}\label{eq_lambda}
\lambda \equiv \frac{\Omega^2} {2\pi G \brho} =
\frac{2\Omega^2 R_1^3}{3GM_1},
\end{equation}
(e.g., \citealt{sto74,cla99}).

Because of tidal and rotational distortions,  the orbital motions
of close binary stars are different from purely Keplerian ones. 
For binaries in spin-orbit synchronization, the angular velocity 
is naturally constrained by equation (\ref{eq_ang_vel}), or its
dimensionless form
\begin{equation} \label{eq_ang_dim}
\Biggl(\frac{\Omega}{\Omega_K}\Biggr)^2 = \frac{4\pi\nu^2}
{\int\Theta^nd^3\xi} \frac{\partial \Psi_g}{\partial \xi} 
\Biggr|_{\boldsymbol{\xi}=(\nu, 0, 0)}.
\end{equation}
However, it turned out in our SCF method 
that $\Omega/\Omega_K$ calculated 
from equation (\ref{eq_ang_dim}) was too small and varied too much to 
give converged results at the very early stage of iterations.
Instead, we empirically found that the condition 
\begin{equation} \label{eq_ang_velocity_dim}
\Biggl(\frac{\Omega} {\Omega_K} \Biggr)^2
=1+\frac{1} {\nu(1-\mu)} \frac{\partial \Psi_g} {\partial \xi} 
\Biggl|_{(\xi,\eta,\zeta)=0},
\end{equation}
corresponding to a vanishingly-small pressure gradient at the center 
of mass of star \I, allows the solutions to converge rapidly,
and is in fact very similar to equation (\ref{eq_ang_dim})
when an equilibrium is achieved. 
In practice, therefore, we first use equation (\ref{eq_ang_velocity_dim}) to 
update $\Omega$ until a temporary convergence is made, 
and then switch to equation (\ref{eq_ang_dim}) to obtain
the desired equilibrium solutions.

To check the accuracy of equilibrium configurations that we 
construct, we use the virial parameter $\epsilon$ defined by 
\begin{eqnarray}\label{eq_virial}
\epsilon=\biggl|\frac{2T+3\int P dV+W} {W} \biggr|,
\end{eqnarray}
where $T=\case{1}{2}\Omega^2M_2A^2\mu^2
-\int\rho\Phi_\Omega d^3x$ is the total kinetic energy and 
$W=\int (\Phi_g+\Phi_{*}) \rho d^3x$ is the total gravitational potential
energy. 

\subsection{Structure Parameters}

A polytrope distorted by tidal and centrifugal forces in general
has different radii in the equatorial and polar directions. 
To measure its size, we use the volume radius 
\begin{equation} \label{eq_re_def}
R_1 = \alpha \xi_1 \equiv \Biggl(\frac{3 V} {4\pi} \Biggr)^{1/3},
\end{equation}
where $V$ denotes the volume enclosed by the isodensity surface with 
$\Theta=0$.  Note that in the absence of perturbing forces, $\xi_1$ is equal
to the first zero of the Lane-Emden functions for unperturbed polytropes
with $n<5$ (e.g., \citealt{cha39}).

Substituting equation (\ref{eq_cden}) into equation (\ref{eq_alpha})
and eliminating $\alpha$ in the resulting equation by use
of equation (\ref{eq_re_def}) yield 
the mass-radius relation 
\begin{equation} \label{eq_K}
K=N_{n}GM_{1}^{(n-1)/n}R_1^{(3-n)/n},
\end{equation}
where $N_n$ denotes the numerical coefficient
\begin{equation} \label{eq_N}
N_n = \frac{4\pi \xi_1^{(n-3)/n}}{(n+1)}
\Biggl(\int \Theta^n d^3\xi\Biggr)^{(1-n)/n}.
\end{equation}
We note that equation (\ref{eq_K}) holds also for unperturbed polytropes, but
with different values of $N_n$.

To characterize the degree of the mass concentration toward the center,
we use 
\begin{equation}\label{eq_concent}
\frac{\rho_c}{\brho} = \frac{4\pi\xi_1^3}{3\int \Theta^n d^3\xi}.
\end{equation}
The dimensionless moment of inertia of star \I\ 
about the $z$-axis is defined by
\begin{equation}\label{moi}
I = \frac{1}{\alpha^5\rho_c}\int \rho  (x^2 + y^2) d^3x.
\end{equation}
Finally, the total angular momentum of star \I\ is given by 
\begin{equation}\label{angm}
J_1 = -\frac{2}{\Omega} \int \rho  \Phi_\Omega d^3x = 
\Jorb + \Jspin,
\end{equation}
where $\Jorb = \Omega M_1 A^2(1-\mu)^2$ and 
$\Jspin = \Omega \int \rho  (x^2 + y^2) d^3x=\alpha^5\rho_c\Omega I$
are the orbital and spin angular momenta of star \I, respectively.
We will study in the next section the dependences of 
$\xi_1$, $N_n$, $\rho_c/\brho$, $I$,
and $\Jspin/\Jorb$ upon the strength of centrifugal and tidal forces. 

\subsection{Expansion of Density-Potential Pair}\label{sec:exp}

For given values of $n$, $\mu$, and $\nu$,
equations (\ref{eq_balance_dim})-(\ref{eq_Pos_dless})
constitute a closed 
set of equations for the equilibrium density $\Theta^n$; 
the rotation parameter $\upsilon$ is related to the 
dimensionless angular velocity $\Omega/\Omega_K$
through equation (\ref{eq_ups_mu_nu}) and
$\Omega/\Omega_K$ is constrained by equation (\ref{eq_ang_dim}) 
for binaries in spin-orbit synchronization.
In an SCF method, the density and the potential are updated alternately 
and iteratively in such a way that at the $i$-th stage of iteration,
equations (\ref{eq_rot_dless})-(\ref{eq_Pos_dless}) are used 
to obtain the potential $\Psi_i$ from $\Theta_i$, which is then 
substituted in equation (\ref{eq_balance_dim}) to yield $\Theta_{i+1}$
for the next iteration stage \citep{ost68}.  

We solve equation (\ref{eq_Pos_dless}) using a mulitpole expansion
technique in the region of space bounded by a sphere with radius $R_L$.
For this, it is convenient to set up spherical coordinates 
$(r, \theta, \varphi)$ centered at the center of mass of star \I.
We divide the sphere into $N_R$ concentric shells and expand the density 
using the associated Legendre functions $P_k^l(\cos\theta)$ as
\begin{equation} \label{eq_rho_expans}
\trho(r_j,\theta,\varphi) =
\sum^{\infty}_{k=0} \sum_{l=0}^{k} \rho_{kl} (r_j) 
P_k^l(\cos\theta) \cos l\phi,
\end{equation}
where $\rho_{kl}$ is the coefficients to be determined, and
\begin{equation}
r_j=\frac{R_L} {\alpha N_{R}} (j-1/2)
\end{equation}
denotes the radius of the $j$-th shell for $j=1,2,\cdots,N_R$.
Note that equation (\ref{eq_rho_expans}) does not contain
$\sin{l\phi}$ terms because of the even symmetry about the  
meridional plane.  Using the orthogonal properties of the Legendre functions,
one can show that 
\begin{equation} \label{eq_rho_approx}
\rho_{kl} (r_j) =\frac{2k+1 } {4\pi} (2-\delta_{l0}) \frac{(k-l)!} {(k+l)!} 
\int\limits_{0}^{\pi} \int\limits_{0}^{2\pi} \trho(r_j,\theta,\varphi) 
P^{l}_{k} (\cos\theta) \cos{l\varphi} \sin{\theta}d\theta d\varphi,
\end{equation}
where $\delta_{l0}$ is the Kronecker delta: 
$\delta_{l0}=1$ for $l=0$; $\delta_{l0}=0$ otherwise
(see e.g., \citealt{hern92}).

For $\Theta^n$ expanded in equation (\ref{eq_rho_expans}),
equation (\ref{eq_Pos_dless}) yields the series solution
\begin{equation} \label{eq_pot_approx}
\Psi_{g} (r,\theta,\varphi) =
- \sum^{\infty}_{k=0} \sum_{l=0}^{k} 
\Bigl[Q_{kl}(r)r^{-k-1} + R_{kl}(r)r^{k} \Bigr]  
P^{l}_{k} (\cos\theta) \cos{l\varphi},
\end{equation}
where
\begin{eqnarray} \label{eq_pot_exp_p1_r}
Q_{kl}(r)=\frac{1} {2k+1} \int\limits_{0}^{r}r'^{k+2} \rho_{kl} (r') dr',
\end{eqnarray}
and
\begin{eqnarray} \label{eq_pot_exp_p2_r}
R_{kl}(r)=\frac{1} {2k+1} \int\limits^{r_{N_R}}_{r}r'^{1-k}\rho_{kl}(r')dr',
\end{eqnarray}
(see e.g., \citealt{bin08}).
Note that equation (\ref{eq_pot_approx}) gives the gravitational potential
at not only the interior but also exterior of a polytrope; 
$\Psi_{g}$ at $r>R_L$ can be obtained by taking 
$Q_{kl}(r)=Q_{kl}(r_{N_R})$ and $R_{kl}(r)=R_{kl}(r_{N_R})=0$.
As noted by \citet{ost68}, the integral representation
via equations (\ref{eq_rho_expans})-(\ref{eq_pot_exp_p2_r}) of the Poisson
equation (\ref{eq_Pos_dless}) is favored over the differential form
(eq.\ [\ref{eq_Pos_dless}]) since the former not only incorporates the
boundary conditions but also guarantees numerical convergence.

\subsection{Iteration Procedure}\label{sec:iter}

To find the self-consistent equilibrium solution $\Theta$ for a 
given set of the parameters $n$, $\mu$, and $\nu$, we proceed the 
following steps. First, as a trial density distribution, 
we consider a uniform sphere with radius $R_L/\alpha$ and density $\Theta=1$,
consisting of $N_R=1200$ shells. 
The sphere is initially rotating at $\Omega/\Omega_K=1$.
Second, we evaluate the integrals in equations 
(\ref{eq_rho_approx}), (\ref{eq_pot_exp_p1_r}) and (\ref{eq_pot_exp_p2_r})  
using Gaussian and Newton-Cotes quadratures 
(e.g., \citealt{pre88})  
to obtain the coefficients $\rho_{kl}(r_j)$, $Q_{ikl}$, and 
$R_{ikl}$ successively.  The corresponding self-gravitational potential 
$\Psi_g$ is then obtained from equation (\ref{eq_pot_approx}) in which 
the summation over the $k$-index
is truncated after 10 terms; the associated error is negligibly small. 
Third, we calculate the rotation parameter $\upsilon$ from
equation (\ref{eq_ups_mu_nu}) and construct the effective potential 
$\Psi$.  Fourth, we solve equations (\ref{eq_balance_dim}) and 
(\ref{eq_ang_velocity_dim}) (or, equation [\ref{eq_ang_dim}] after
a temporary convergence) to update $\Theta$ and $\Omega/\Omega_K$.  
Fifth, we calculate the center of mass 
$\boldsymbol\xi_{\rm cm} = (\int \boldsymbol\xi \Theta^n d^3\xi)/
( \int\Theta^n d^3\xi) $ of star \I\ using the updated $\Theta$, and
shift the coordinates so as to make $\boldsymbol\xi_{\rm cm}$ coincide with 
the origin, while keeping star \II\ at $\boldsymbol\xi=(\nu, 0, 0)$ 
in the new coordinates. 
As a convergence criterion, we require the virial parameter 
$\epsilon$ from equation (\ref{eq_virial}) to be less than $10^{-10}$; 
otherwise, we go back to the second step with the updated $\Theta$ and  
$\Omega/\Omega_K$ in the shifted coordinates, and 
repeat the iterations until the convergence is attained.
For stars with critical rotation, we find that
typically 50 iterations suffice to lead to converged results.

\section{Numerical Results} \label{sec_numerical_results}

While the main purpose of this work is to find the changes in the internal
structure of binary stars due to rotational and tidal perturbations, 
we first apply our method to construct equilibrium models of polytropes
in isolation or in uniform rotation.  This will allow us to check
our SCF method as well as to quantify the effects of rotation on the 
stellar structure.  We then explore equilibrium polytropes subject to both
centrifugal and tidal forces.

\subsection{Undisturbed Polytropes}

We first consider undisturbed polytropes, the solutions to
which can be obtained by taking $\upsilon=0$, corresponding 
to $\Psi_\Omega=\Psi_*=0$, in our SCF method.
Table \ref{ta_unperturbed_comparison} lists the resulting 
dimensionless parameters that we find for some selected values of $n$.  
All the values are in excellent agreement with those given in 
\citet{cha39}, confirming the performance of our technique.

\subsection{Uniformly Rotating Polytropes}\label{sec_rot}

\begin{figure}
\epsscale{1.}
\plotone{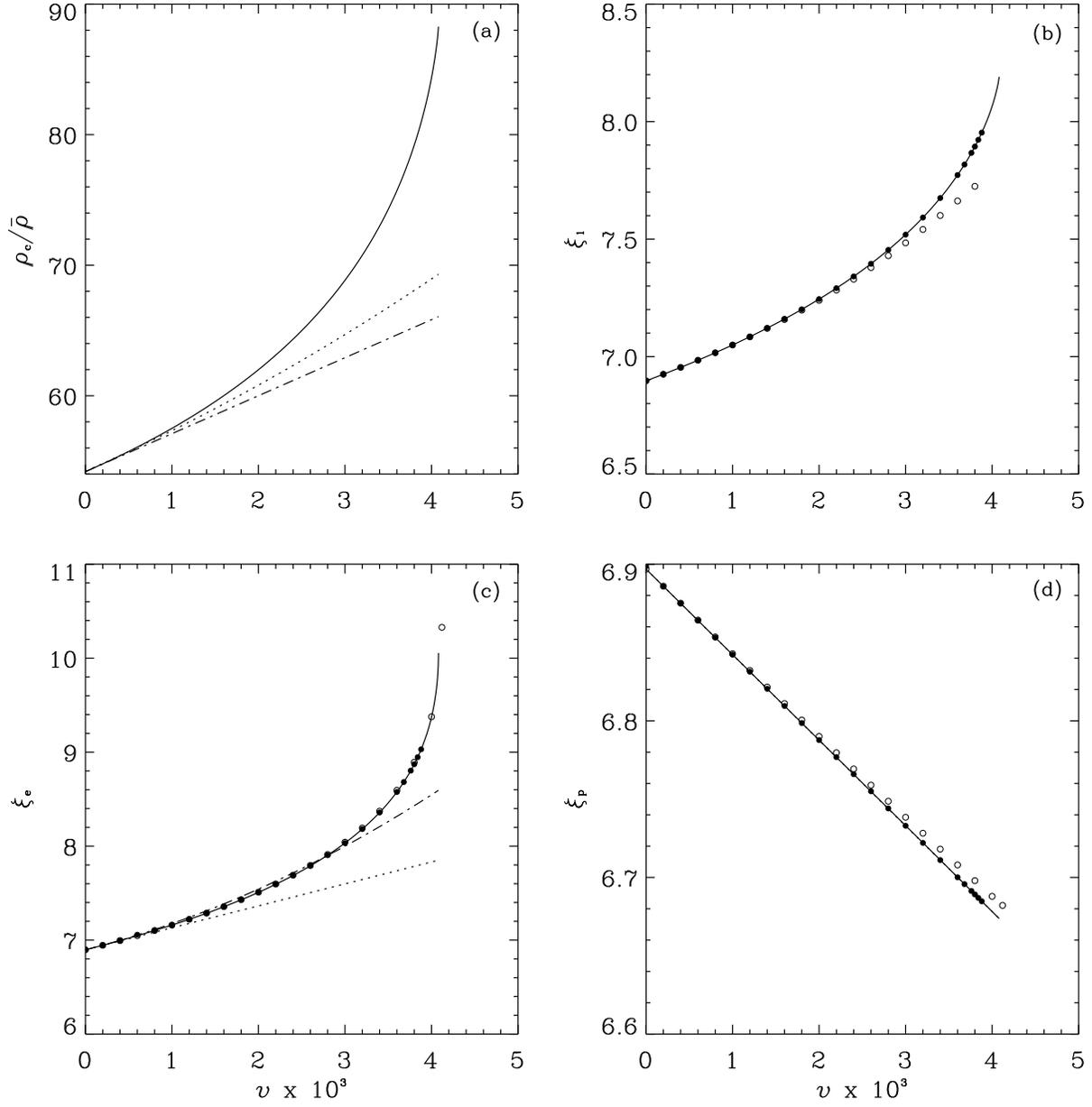}
\caption{ Dependences on the rotation parameter $\upsilon$  
of (\textit{a}) the mass concentration $\rho_c/\brho$, 
(\textit{b}) volume radius $\xi_1$,
(\textit{c}) equatorial radius $\xi_e$,
and (\textit{d}) polar radius $\xi_p$ 
for rotating polytropes with $n=3$.
Solid lines plot the results of the present study,
while dotted and dot-dashed lines are for the first- and
second-order perturbation analyses of \citet{cha33a} and 
\citet{ana68}, respectively.  Open circles denote the values from 
the advanced perturbation theory of \citet{lin77, lin81}, 
while the direct numerical results of
\citet{jam64} are given by filled circles. 
}\label{pic_rot1}
\end{figure}

\begin{figure}
\epsscale{1.}
\plotone{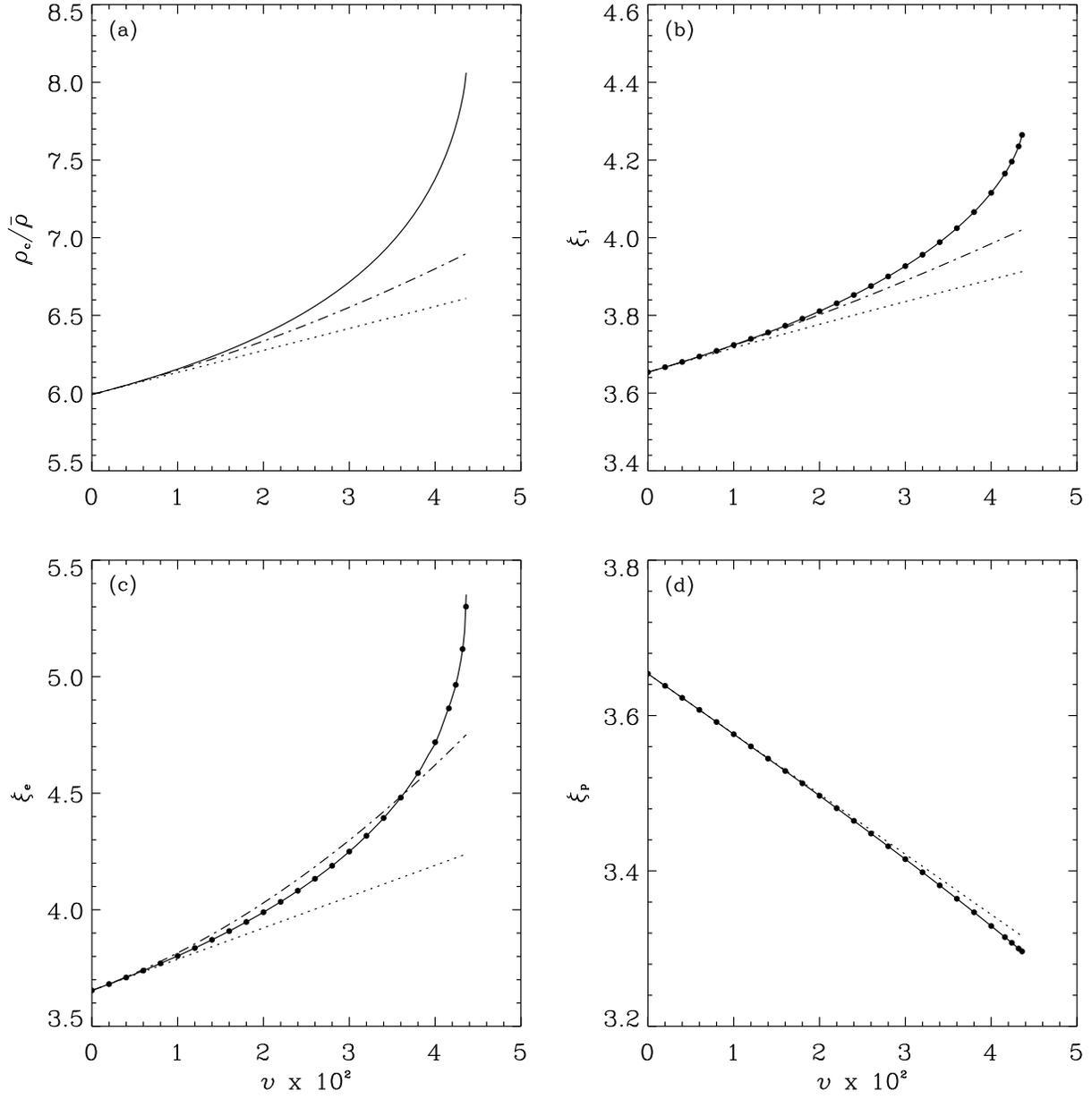}
\caption{
Same as Figure \ref{pic_rot1} except for $n=1.5$ polytropes.
} \label{pic_rot2}
\end{figure}

If one takes $\mu=1$ in equations (\ref{eq_rot_dless}) and 
(\ref{eq_ext_dless}) and regards $\upsilon$ as a free parameter,
the problem is reduced to finding solutions for polytropes under
uniform rotation.  
Figures \ref{pic_rot1} and \ref{pic_rot2} give as solid lines 
the dependences upon the rotation parameter $\upsilon$ of 
the mass concentration $\rho_c/\brho$,  
dimensionless volume radius $\xi_1$, polar radius
$\xi_p=R_p/\alpha$, and equatorial radius $\xi_e=R_e/\alpha$
for rotating polytropes with $n=3$ and $1.5$, respectively.
For comparison, we plot the results available from the published work, 
based on the first-order perturbation theory \citep{cha33a},
the second-order approximation \citep{ana68},
the advanced first-order approximation \citep{lin77,lin81},
and the Legendre-polynomial expansions \citep{jam64}
as dotted lines, dot-dashed lines, open circles, and filled circles,
respectively.  

As $\upsilon$ increases from zero, the centrifugal force tends to distort
the shape of a polytrope by increasing the equatorial radius 
at the expense of the polar radius. 
The polar compression is smaller than the equatorial expansion, 
causing the polytrope to have a larger mean radius than the 
non-rotating counterpart (e.g., \citealt{lin81}).  
A polytrope with faster rotation should definitely be more centrally 
condensed in order for self-gravity to offset the enhanced centrifugal 
and pressure gradient forces near the equatorial plane, as 
Figures \ref{pic_rot1}\textit{a} and \ref{pic_rot2}\textit{a} evidence.
When $\upsilon$ becomes sufficiently large, perturbed polytropes reach 
a critical state where they are on the verge of equatorial breakup.
This critical rotation is formally expressed as 
$\partial \Phi/\partial r=0$ at the equator,
or equivalently,
$\partial \Theta/\partial \xi|_{\xi =\xi_e}=0$ when $\upsilon=\upsilon_c$,
indicating that the centrifugal force barely balances
the self-gravity at the equator.  
No equilibrium configuration exists for $\upsilon>\upsilon_c$.
Table \ref{ta_rot_terminated_values} gives the various parameters 
for the critically-distorted uniformly-rotating 
polytropes with differing $n$.  The smaller $n$, the smaller 
$\rho_c/\brho$ and the larger deviation of the gravitational potential 
of star \I\ from that of a Newtonian point-mass.  
The values of $\lambda_c$ for polytropes with indices $n=1.0-2.5$
agree, within less than 0.2\%, with the published numerical results of 
\citet{jam62} (see also \citealt{hur64}).
Note that $\lambda_c=0.36074$ for $n=4.5$ polytropes are virtually
the same as the value from the Roche model with $\mu=1$ \citep{hor04}.

Although the first- and second-order perturbation methods are
reasonably good for $\xi_p$ since the centrifugal force is very small 
near the poles, they systematically underestimate 
$\rho_c/\brho$, $\xi_e$, and $\xi_1$ as $\upsilon$ increases toward
$\upsilon_c$.  Being valid only for $\upsilon\ll1$, 
they are also unable to yield correct values of $\upsilon_c$. 
On the other hand, for the the whole range of $\upsilon$,
our results for $\xi_p$ and $\xi_e$ are within 1\% of the 
numerical values from \citet{jam64}, 
verifying again the accuracy of our SCF method.
For practical purposes,  we fit the dependences upon $\upsilon/\upsilon_c$ 
of the various parameters characterizing the shapes and internal 
structures of rotating polytropes 
using a simple function 
\begin{equation}\label{eq_rot_fit}
f(\upsilon)=a_0+a_1\Bigl(\frac{\upsilon}{\upsilon_c}\Bigr)
   +a_2\Bigl(\frac{\upsilon}{\upsilon_c}\Bigr)^2
   +a_3\ln{\Bigl(1-a_4\Bigl[\frac{\upsilon}{\upsilon_c}\Bigr]\Bigr)},
\end{equation}
with coefficients $a_i$.  
Table \ref{tab_rot_fit} lists 
the fitting coefficients for $f=\xi_1$, $\xi_e$, $\xi_p$, $\rho_c/\brho$,
$N_n$, and $I$ of rotating polytropes with $n=1.5$ and $n=3$; 
the associated fitting errors are less than 1\%.
Note that $a_0$ in each fit is identical to the 
corresponding value of non-rotating polytropes.
 
It is interesting to compare the shapes of rotationally-distorted polytropes 
with those under the Roche approximation that assumes that the stellar gravity
is dominated by the central mass concentration and that the polar radius 
is unchanged due to rotation.
The main predictions of the Roche models are that
the ratio of the equatorial velocity $\vel_e$ to
the critical value $\vel_c$ behaves with the aspect ratio
$\mathcal{R}\equiv \xi_e/\xi_p$ as
\begin{equation}\label{eq_vel_ratio}
\frac{\vel_e}{\vel_c} = \sqrt{\frac{3(\mathcal{R}-1)}{\mathcal{R}}},
\end{equation}
and that $\mathcal{R}_c=1.5$ in the case of critical rotation
(e.g., \citealt{col63,bau99,tow04}).
Figure \ref{pic_crit}\textit{a} plots as a solid line the change 
in $\mathcal{R}_c$ of rotating polytropes with varying $n$, while
Figure \ref{pic_crit}\textit{b} gives $\vel_e/\vel_c$ as a function of
$\mathcal{R}$ for various polytropic and the Roche models.
In terms of $\mathcal{R}_c$, the Roche models are a good approximation 
for polytropes with very high mass concentration toward their centers
(relative errors less than 3\% for $n \simgt 2.0$).
On the other hand, 
equation (\ref{eq_vel_ratio}) is quite accurate for polytropes with $n=1.5$ 
except near the critical configuration. It underestimates 
$\vel_e/\vel_c$ for polytropes with $n\simgt 3.0$ by about $10\%$,
while overestimating for $n=0.5$ polytropes by about $15\%$.
This suggests that one should be cautious when 
estimating the equatorial velocity from an observed
aspect ratio of a rapidly rotating star. 

\subsection{Rotationally and Tidally Distorted Polytropes}

\begin{figure}
\epsscale{1.}
\plotone{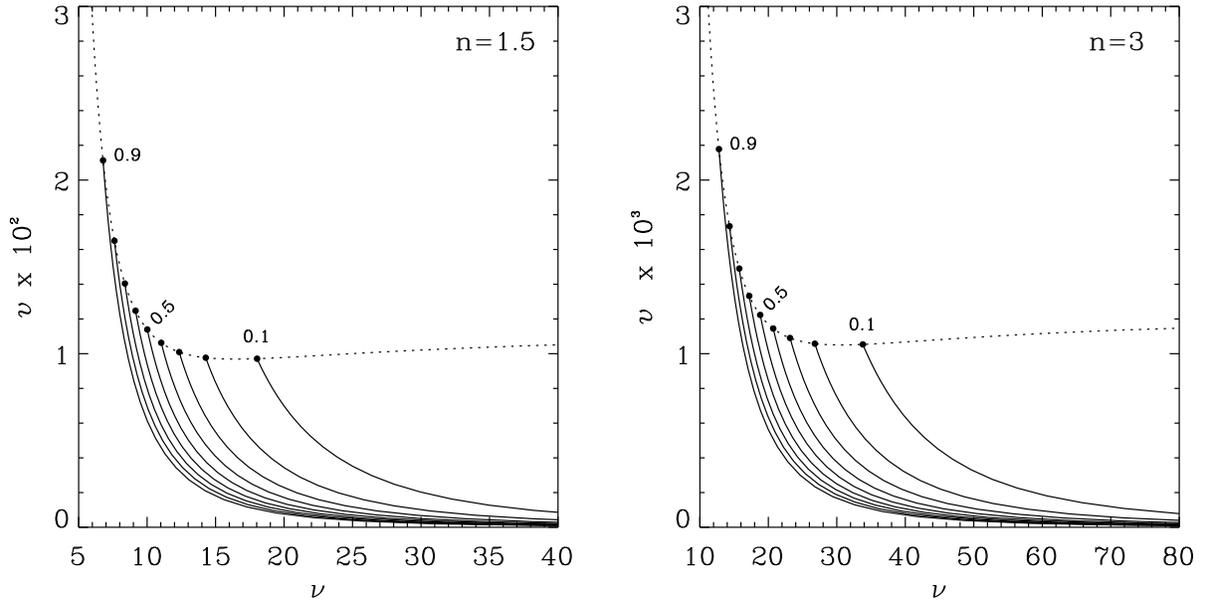}
\caption{Dependences of the rotation parameter $\upsilon$
on the dimensionless orbital separation $\nu$ for self-consistent 
equilibrium polytropes with (\textit{left}) $n=1.5$ and 
(\textit{right}) $n=3$ under both rotational 
and tidal perturbations. 
Each solid line corresponds to a sequence of equilibrium polytropes
with fixed relative mass $\mu$ indicated at one end of the line.
The critical configuration at each sequence is marked by a filled circle.
The dotted line connecting the filled circles demarcates the boundary
in the $\nu$-$\upsilon$ plane above which no equilibrium polytrope exists.
}\label{pic_ups_nu}
\end{figure}

When a binary star with a given polytropic index $n$ is subject to both tidal 
and centrifugal distortions, 
its internal structure is determined solely by the two parameters in our
formulation: 
its relative mass $\mu$ and the dimensionless orbital separation $\nu$; 
for systems in synchronous rotation, 
the rotation parameter $\upsilon$ is given automatically
via equations (\ref{eq_ups_mu_nu}) and (\ref{eq_ang_dim}). 
Figure \ref{pic_ups_nu} plots the dependences of $\upsilon$
upon $\nu$ for the equilibrium polytropes under both tidal and
centrifugal forces.  Each solid line corresponds to a
sequence of equilibrium models with a fixed relative mass $\mu$ 
displayed in the panels.
Note that $\upsilon$ monotonically decreases with increasing orbital 
separation along each sequence, becoming vanishingly 
small at $\nu\rightarrow\infty$.
On the other hand, a polytrope becomes distorted more and more 
as $\nu$ decreases along a given sequence. 
When $\nu$ becomes sufficiently small,
it eventually reaches a critical state beyond which self-gravity no longer 
balances the combined tidal and centrifugal forces at the equator.
The filled circles marked at the uppermost tips of the sequences
correspond to the critical points $(\nu_c,\upsilon_{c})$.
In what follows, we use the term ``critical component'' to refer to 
star \I\ in critical configuration.  The dotted line connecting the 
critical points draws the locus of the critical components in 
the $\nu$-$\upsilon$ plane, above which no equilibrium 
polytrope exists.

\subsubsection{Shape, Size, and Internal Structure}\label{structure}

\begin{figure}[c!]
\epsscale{1.00}
\plotone{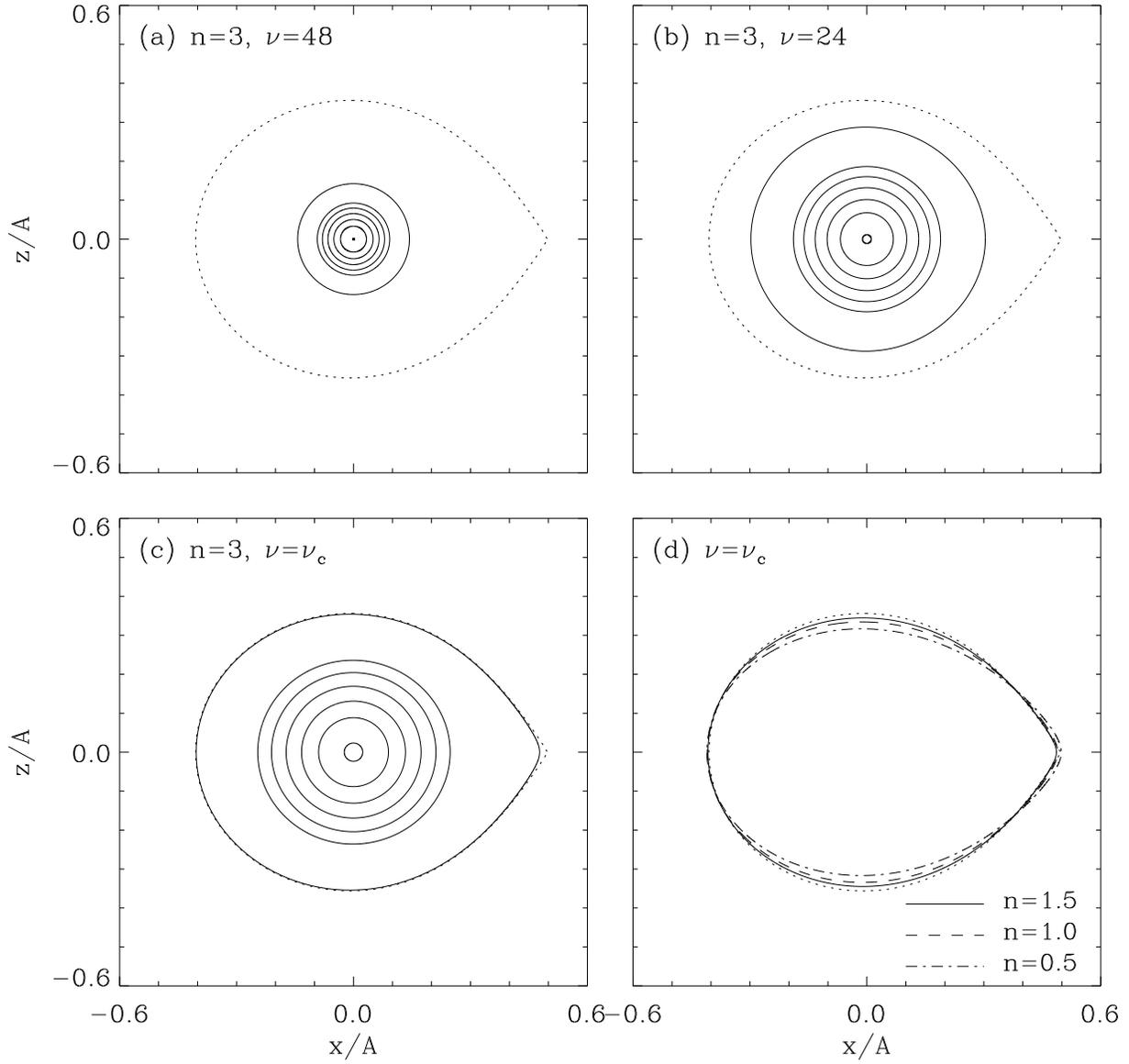}
 \caption{\textit{a}-\textit{c}: Meridional cross-sections of constant 
density surfaces of distorted equilibrium polytropes with $n=3$. 
Solid contours correspond to $\rho/\brho=3^{-m} (\rho_c/\brho)_{\rm iso}$ 
with $m=0,1,\cdots,5,\infty$ from inside to outside. 
\textit{d}: Outer boundaries 
of the distorted polytopes in critical configuration.
In all the panels, the dotted line draws the Roche lobe around star \I\
that has a half of the total mass ($\mu=0.5$).
See text for details.
}\label{pic_shape}
\end{figure}

Figure \ref{pic_shape}\textit{a-c} illustrates how the shapes 
on the meridional plane of $n=3$ equilibrium polytropes 
change as the amplitude of tidal and rotational perturbations varies.
All the models have $\mu=0.5$.  The corresponding Roche lobe around 
star \I\ is plotted as a dotted line.
In each panel, the inner solid contours draw equidensity surfaces 
with $\rho/\brho=3^{-m} (\rho_c/\brho)_{\rm iso}$,
where $m=0, 1,\cdots,5$ from inside to outside, 
while the outermost contour represents the boundary of a 
perturbed polytrope.  Here and in what follows, the subscript ``iso'' 
indicates the quantities of the corresponding isolated polytrope.
When $\nu$ is sufficiently large, the tidal and rotational perturbations
are so weak that a polytrope is in an almost spherical configuration,  
with the central density only slightly larger than the unperturbed value
(Fig.\ \ref{pic_shape}\textit{a}). 
As $\nu$ decreases, the enhanced perturbations make the polytrope larger 
and more centrally concentrated (Fig.\ \ref{pic_shape}\textit{b}).
It is remarkable that when $\nu=\nu_c$ ($=18.81$ for $n=3$), 
the outer envelope 
of the critical component fills the Roche lobe almost completely, 
while the dense inner-part still remains nearly spherical 
(Fig.\ \ref{pic_shape}\textit{c}).

\begin{figure}
\epsscale{1.00}
\plotone{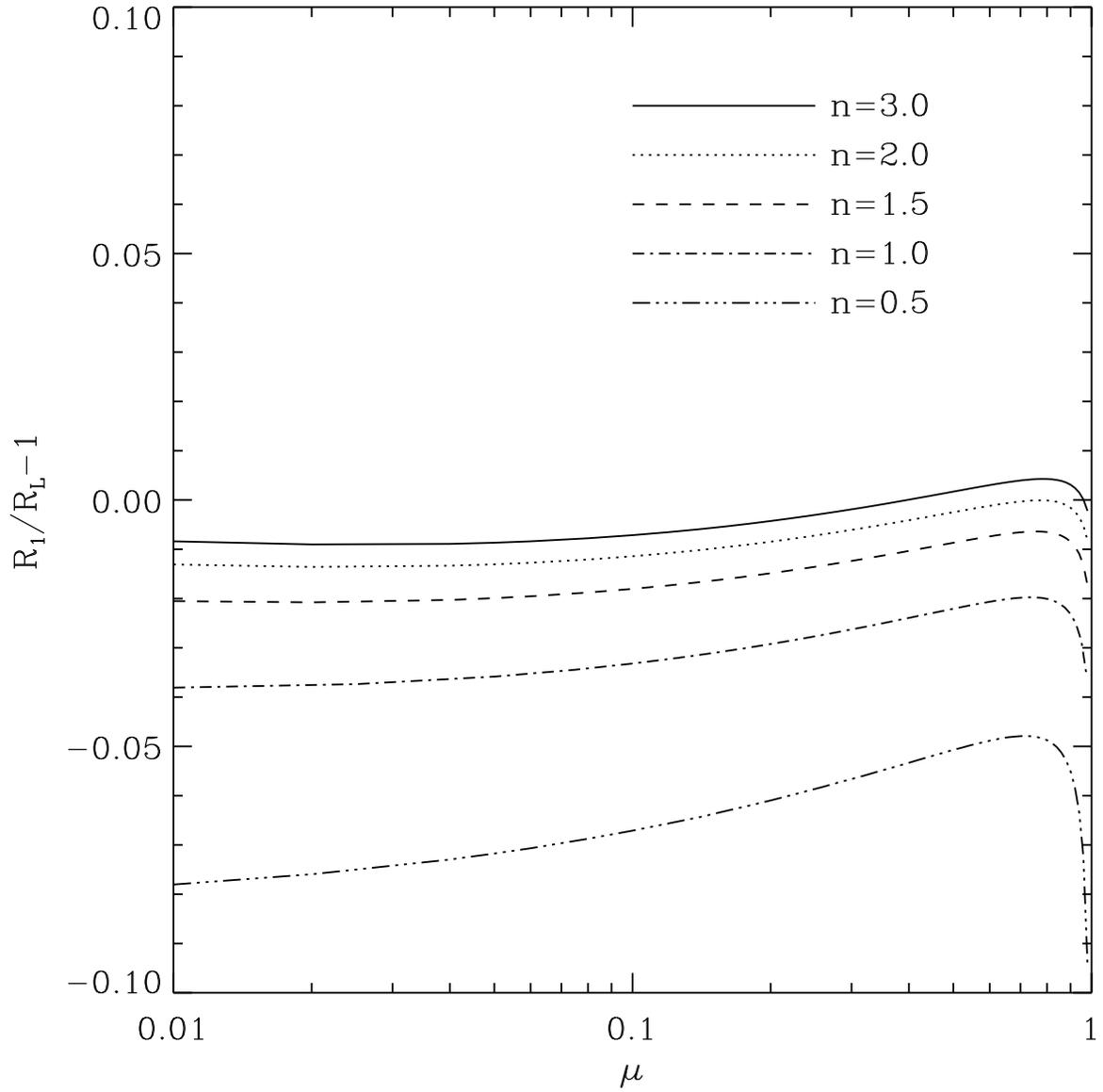}
 \caption{Comparison of the volume radius $R_1$ of a critical component
with the effective radius $R_L$ of the corresponding Roche model. 
}\label{pic_Roche}
\end{figure}

Figure \ref{pic_shape}\textit{d} plots the outer boundaries of
the critical components with differing $n$ for $\mu=0.5$.
When $n\simgt 2$, a critical component has a boundary
virtually identical to its Roche lobe.   
With relatively low central concentrations, 
polytropes with $n\simlt 1$ become flattened toward the equatorial plane, 
although the deviation from the Roche lobe is still slight.
Figure \ref{pic_Roche} gives a quantitative comparison between 
the volume radius $R_1$ of a critical component and 
the effective radius $R_L$ of the Roche lobe fitted by \citet{egg83}  
\begin{equation}\label{eq_eggleton}
\frac{R_L}{A}=\frac{0.49q^{2/3}} {0.6q^{2/3}+\ln(1+q^{1/3})},
\end{equation}
where $q\equiv M_1/M_2=\mu/(1-\mu)$.
Indeed, Eggleton's formula is an excellent approximation to $R_1$, 
with a relative error less than 2\%, for polytropes with $n\simgt 2$. 
For small values of $n$, $R_L$ overestimates $R_1$
because of the equatorial flattening.  For $n=3/2$ polytropes, we find
\begin{equation}\label{eq_r_over_A}
\frac{R_1}{A}=\frac{0.5126q^{0.7388}} 
{0.6710q^{0.7349}+\ln(1+q^{0.3983})},\;\;\;{\rm for}\;\; n=3/2,
\end{equation}
gives a good fit (within $\sim1-2\%$ for $10^{-2}<q<10^{4}$)
to the volume radius.

\begin{figure}
\epsscale{0.90}
\plotone{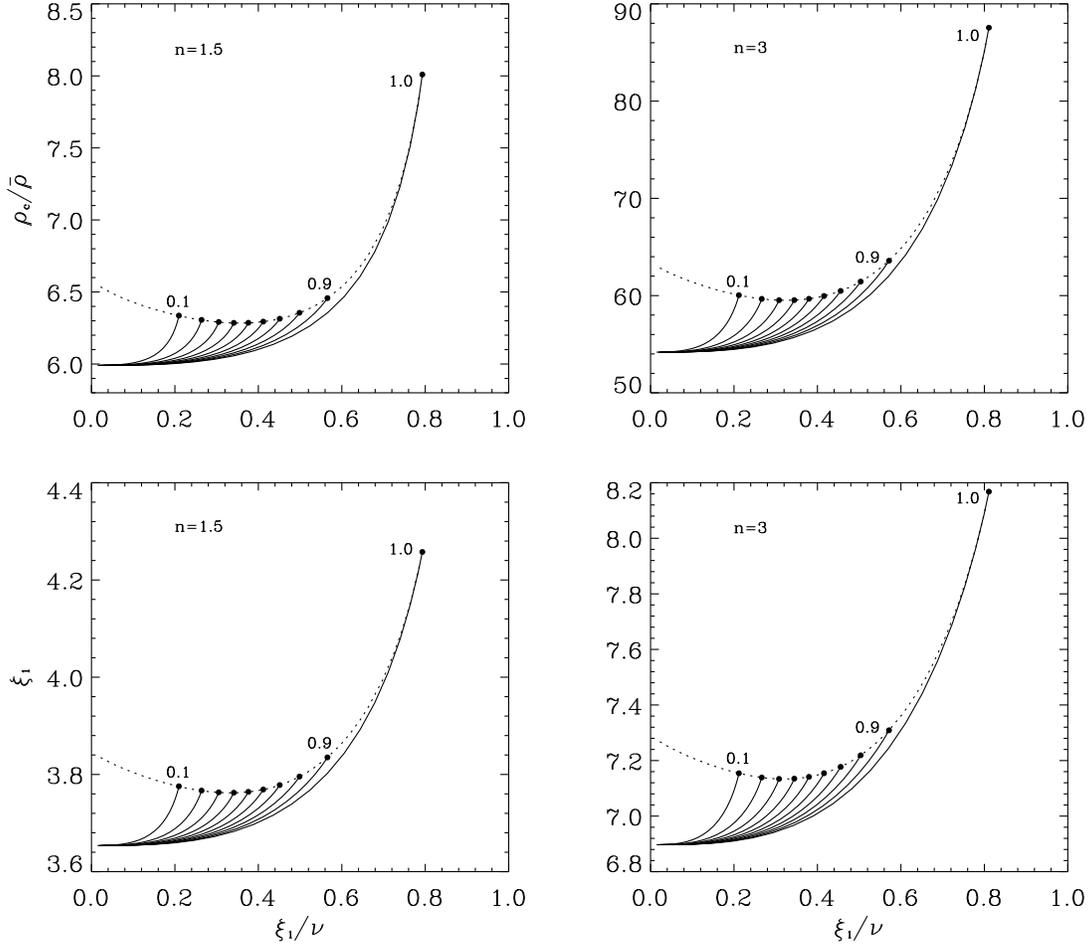}
\caption{
Dependences of $\rho_c/\brho$ and $\xi_1$
on the relative radius $R_1/A=\xi_1/\nu$ for 
(\textit{left}) $n=1.5$ and  (\textit{right}) $n=3.0$ polytropes.
Each solid line is a sequence of distorted polytropes with 
fixed relative mass $\mu$ indicated at one end of the line. 
The dotted lines connecting the filled circles, denoting 
the end points of the sequences,  correspond to the critical components.
}\label{pic_ksi_ra}
\end{figure}

\begin{figure}
\epsscale{0.90}
\plotone{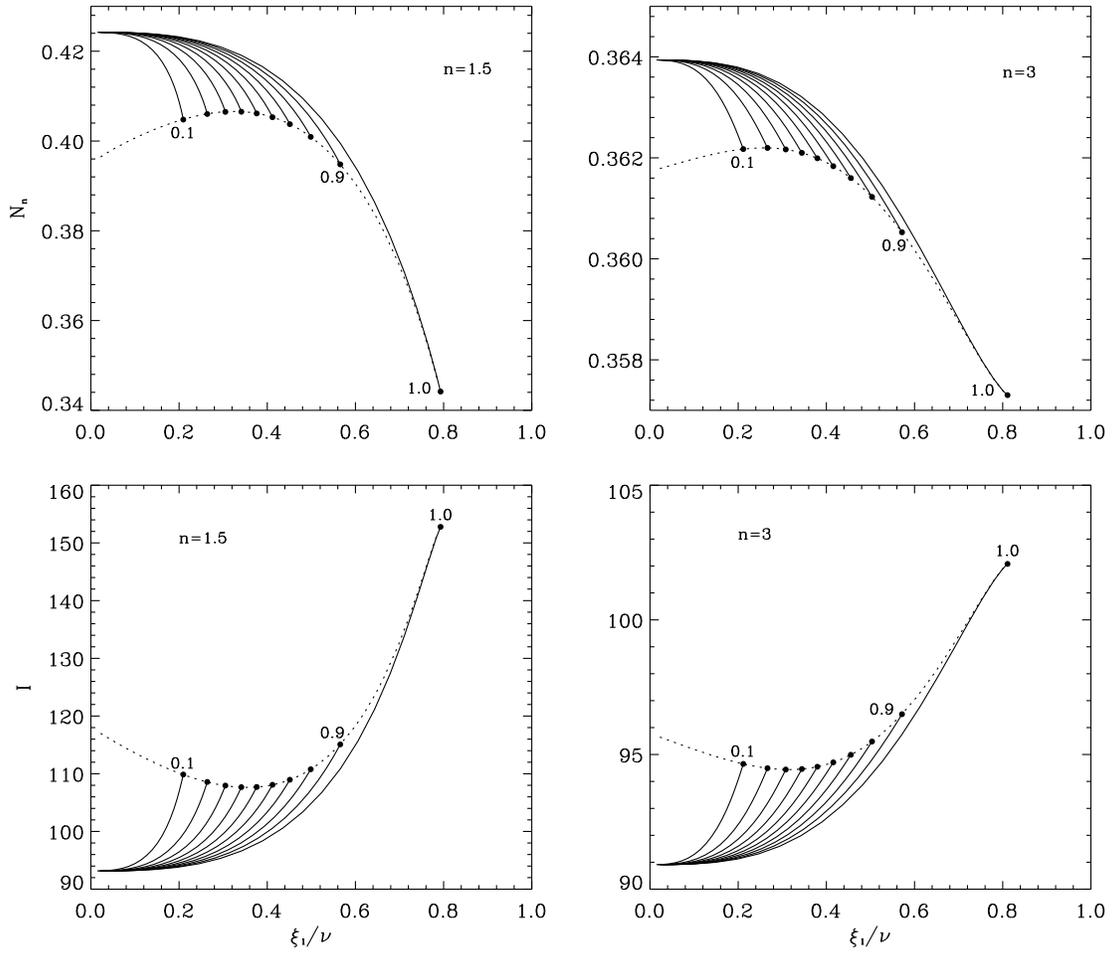}
\caption{
Same as Figure \ref{pic_ksi_ra} for $N_n$ and $I$.
}\label{pic_I_ra}
\end{figure}

Figures \ref{pic_ksi_ra} and \ref{pic_I_ra} plot the variations of 
$\rho_c/\brho$, $\xi_1$, $N_n$, and $I$ of 
rotationally- and tidally-distorted polytropes with
$n=1.5$ and 3.0 against the relative radius $R_A\equiv 
\xi_1/\nu=R_1/A$ as solid lines.
Again, the dotted line connecting the top ends of the constant-$\mu$ sequences
in each panel corresponds to the critical components.  As expected,
the central density concentration, volume radius, and moment of inertia 
increase as the perturbation amplitude increases. 
The relative increment of
$\rho_c/\brho$ is greater for polytropes with larger $n$.
Note that the largest changes in these quantities occur when $\mu=1$,
corresponding to purely rotating polytropes;  in synchronized
binaries with $0.1\simlt \mu \simlt 0.9$, the variations of $\rho_c/\brho$,
$\xi_1$, and $N_n$ are less than 10\% even in critical configurations.
The dependences of these quantities on $R_A$
for a fixed value of $\mu$ can be approximated by
\begin{equation} \label{eq_ra_fit}
g(R_A)=b_0+b_1R_A+b_2R_A^2+b_3R_A^3~\exp{(b_4 R_A)},
\end{equation}
with $g=\rho_c/\brho$, $\xi_1$, $N_n$, or $I$.  The fitting 
coefficients $b_i$ are given in Tables \ref{tab_rho_app},
\ref{tab_xi_app}, \ref{tab_Nn_app}, and \ref{tab_I_app}, 
respectively; the fits are 
accurate within $1\%$. 
Note that $b_0$'s in each Table are identical to the values for 
the unperturbed polytropes.

\begin{figure}
\epsscale{1.00}
\plotone{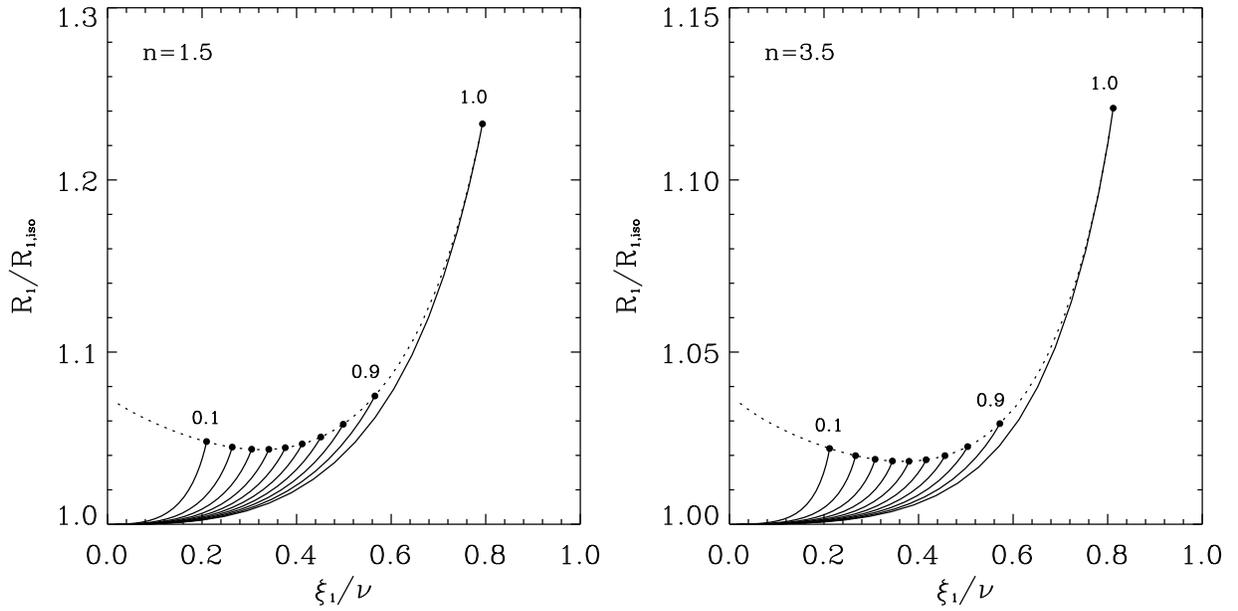}
\caption{Changes in the radius of a star with fixed mass due to tidal and 
rotational perturbations under the assumption of constant pressure constant 
$K$ for (\textit{left}) $n=1.5$ and (\textit{right}) $n=3.5$ polytropes.
Each solid line is a sequence of distorted polytropes with
fixed relative mass $\mu$ indicated at one end of the line.
The dotted lines correspond to the critical components.
}\label{pic_ee_ra}
\end{figure}

Since $\alpha$ depends on $\rho_c$ which in turn varies with $\nu$ and 
$\mu$, the dimensionless volume radius $\xi_1$ shown in 
Figure \ref{pic_ksi_ra}  does not give direct 
information on the change of a stellar size due to centrifugal and 
tidal forces.  Assuming that the perturbations do not modify the pressure 
constant $K$ (equivalently, specific entropy or degeneracy;
see e.g., \citealt{bur93}) inside a star, 
equation (\ref{eq_K}) gives 
$R_1/R_{1, \rm iso} = (N_n / N_{n, \rm iso})^{n/(n-3)}$
as the size ratio of distorted to unperturbed polytopes with the same mass.
Figure \ref{pic_ee_ra} plots $R_1/R_{1, \rm iso}$ as solid lines 
for polytrope with $n=1.5$ and $3.5$, showing that
stars become bigger as the amplitude of perturbations increases.\footnote{As 
is well known, it is not viable to constrain the radius from the mass for
polytropes with $n=3$ (e.g., \citealt{cha39}).}, 
For $n=3.5$ polytropes in critical configuration, the enhancement factor 
in the stellar size
is $\sim12\%$ for $\mu=1$ and $\sim 2-3\%$ for $\mu\sim0.1-0.9$, while
it increases to $\sim23\%$ for $\mu=1$ and $\sim 4-8\%$ for $\mu\sim0.1-0.9$
for $n=1.5$ polytropes. 
Taken together with the results for $\rho_c/\brho$ 
discussed above, this implies that polytropes with smaller $n$ adjust
themselves to the perturbations by increasing their size more than 
being centrally condensed, while those with larger $n$ do so by 
increasing the density concentration more.

\begin{figure}
\epsscale{1.00}
\plotone{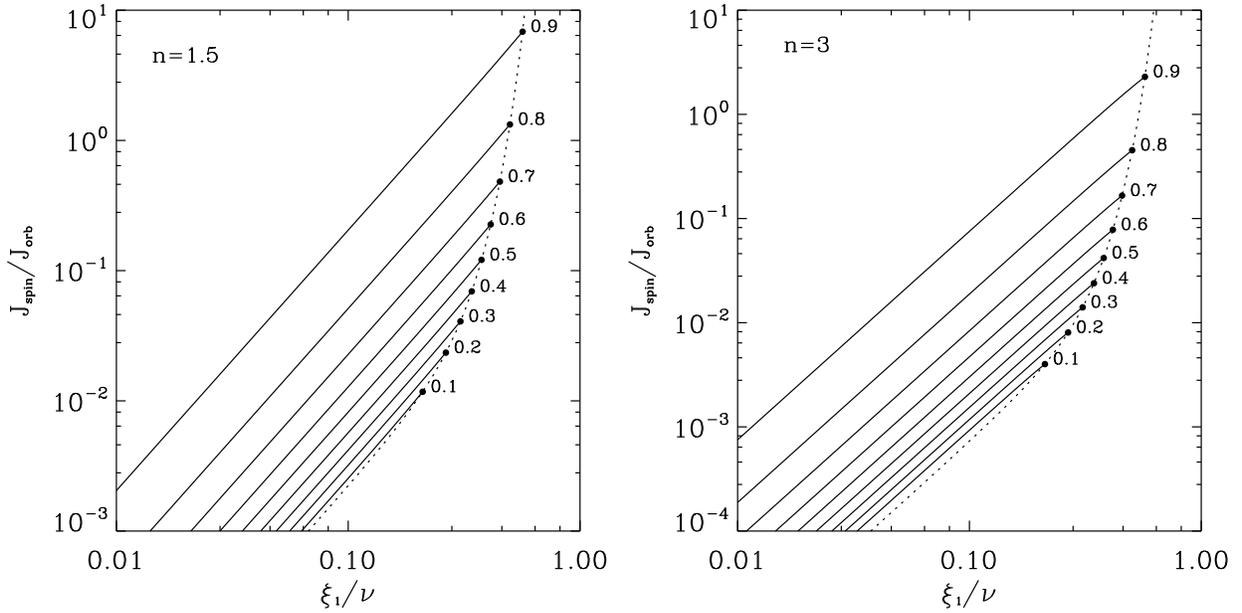}
\caption{
Ratio of the spin to orbital angular momenta, $\Jspin/\Jorb$, 
of (\textit{left}) $n=1.5$ and (\textit{right}) $n=3.0$ polytropes, 
as functions of $\xi_1/\nu$ and $\mu$.
Each solid line corresponds to a sequence of distorted polytropes with 
fixed relative mass $\mu$ indicated at one end of the line.
The dotted lines connecting the filled circles, denoting 
the end points of the sequences,  represent the critical components.
}\label{pic_jj}
\end{figure}

In the dynamical study of close binary systems such as mass
transfer, the spin angular momentum is often neglected compared to the 
orbital angular momentum.  To check if this is a reasonable assumption,
we plot in Figure \ref{pic_jj} the dependences of $\Jspin/\Jorb$
on $\xi_1/\nu$ and $\mu$ as solid lines.  Overall, 
$\Jspin/\Jorb$ is well approximated by 
\begin{equation} \label{eq_jj_fit}
\frac{\Jspin}{\Jorb}= \frac{c_0}{(1-\mu)^2}
\Biggr(\frac{\xi_1}{\nu}\Biggr)^2,
\end{equation}
with the coefficient $c_0=0.20460$ and 0.07536 for $n=1.5$ and 3 cases,
respectively.\footnote{In the limit of $\xi_1/\nu\rightarrow0$, 
it can be shown that $c_0=3I(\rho_c/\brho)(4\pi\xi_1^5)^{-1}$ for
unperturbed polytropes.}  
The fit is almost exact for small $\xi_1/\nu$, and
the maximum error occurring at the critical values of $\xi_1/\nu$
is less than 5\%.
Apparently, the contribution of the spin angular momentum to the total is 
negligible for wide binaries where $\xi_1/\nu\ll1$.  
Notice, however, that when a binary star with sufficiently large $\mu$ 
is near the critical configuration, 
$\Jspin$ even exceed $\Jorb$ and can thus be dynamically important.

For the critical components, $\rho_{c}/\brho$,  
$N_n$, $\Jspin/\Jorb$, and $I$ can be approximated within 1\% of accuracy by
\begin{mathletters}\label{eq_rhoc_fit}
\begin{eqnarray}
\rho_{c}/\brho& = &5.9907+ \frac{0.2755q^{0.8849}}
{0.2430q^{0.8258}+\ln(1+q^{1.1415}) },\;\;{\rm for}\;\;
n=1.5,\\
\rho_{c}/\brho& = &54.1825+\frac{5.3141q^{0.9486}}{0.2712
q^{0.8933}+\ln(1+q^{1.1494}) },\;\;{\rm for}\;\; n=3,
\end{eqnarray}
\end{mathletters}
\begin{mathletters}\label{eq_Nn_fit}
\begin{eqnarray}
N_n&=&0.42422-\frac{0.0191q^{0.9561}}
{0.3557q^{0.9130}+\ln(1+q^{1.1635})},
\;\;{\rm for}\;\;n=1.5,\\
N_n&=&0.36394-\frac{0.00207q^{1.01714}}
{0.35430q^{1.00296}+\ln(1+\mu^{1.08931})}, \;\;{\rm for}\;\;n=3,
\end{eqnarray}
\end{mathletters}
\begin{mathletters}\label{eq_jj_crit}
\begin{eqnarray}
\frac{\Jspin}{\Jorb} &=&
\begin{cases}  
0.076q^{1.858}+0.043q^{0.645},       & \text{if $q<0.5$}, \\
0.068(q-0.5)^{2.106}+0.100q^{0.9},  &\text{if  $q\geq 0.5$.},
\end{cases} 
\;\;{\rm for}\;\;n=1.5, \\
\frac{\Jspin}{\Jorb} &=&
\begin{cases}  
0.026q^{1.853}+0.015q^{0.651},     & \text{if $q<0.5$}, \\
0.026(q-0.5)^{2.061}+0.037q^{0.9}, & \text{if $q\geq 0.5$,}
\end{cases}
\;\;{\rm for}\;\;n=3, 
\end{eqnarray}
\end{mathletters}
and 
\begin{mathletters}\label{eq_I_crit}
\begin{eqnarray}
I&=&93.1560+\frac{15.5476q^{0.9334}}{0.3795q^{0.8902}+\ln(1+q^{1.1985})},
\;\;{\rm for}\;\;n=1.5, \\
I&=&90.910+\frac{4.4373q^{0.9879}}{0.5200q^{0.9560}+\ln(1+q^{1.1593})},
\;\;{\rm for}\;\;n=3,
\end{eqnarray}
\end{mathletters}
for the range of $0.05<q<40$.

\subsubsection{Rotational vs.\ Tidal Distortion}

\begin{figure*}[c!]
\epsscale{1.00}
\plotone{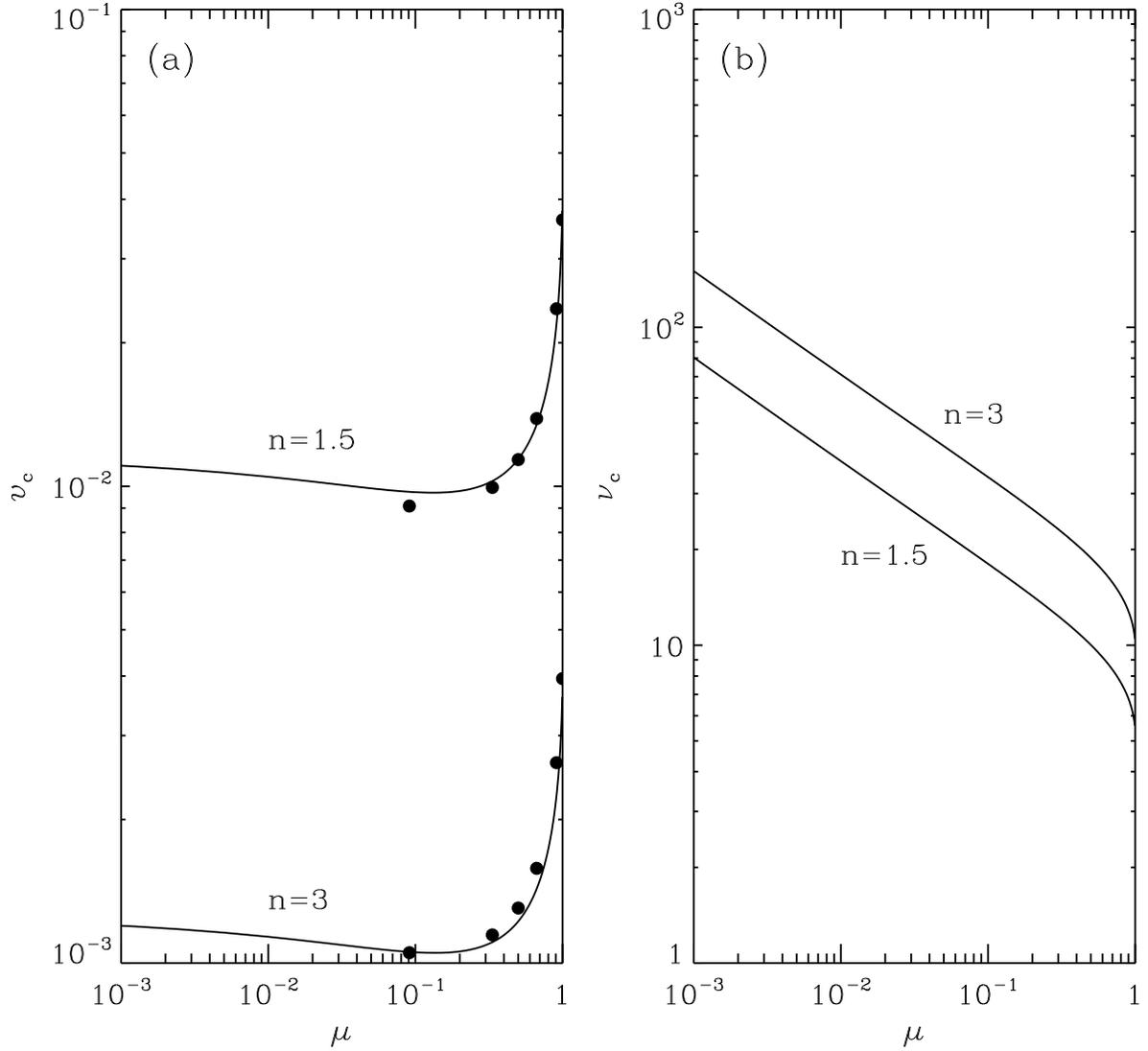}
\caption{
Dependences on $\mu$ of 
(\textit{a}) the critical rotation parameter $\upsilon_c$ and 
(\textit{b}) the critical dimensionless orbital separation $\nu_c$ 
for polytropes with $n=1.5$ and 3.0.  In (\textit{a}),
the filled circles plots the results of \citet{sin83} based on
the double approximation method.
}\label{pic_nu_ups_mu}
\end{figure*}

Figure \ref{pic_nu_ups_mu} plots as solid curves $\upsilon_{c}$ and 
$\nu_{c}$ for the critical components with $n=1.5$ and $3.0$ 
as functions of $\mu$.  These are fitted to
\begin{mathletters}\label{eq_upsilonc_fit}
\begin{eqnarray}
\upsilon_{c}\cdot10^3&=&4.08028+4.10(1-\mu)-0.54(1-\mu)^{2}-6.58(1-\mu)^{0.46}
\;\;\text{for}\;n=1.5, \\
\upsilon_{c}\cdot10^2&=&4.35828+4.55(1-\mu)-0.79(1-\mu)^{2}-7.15(1-\mu)^{0.43}
\;\;\text{for}\;n=3, 
\end{eqnarray}
\end{mathletters}
and
\begin{equation}\label{eq_crit_nu}
\nu_c= 
\begin{cases}
\sum_{i=0}^5 d_i (1-\mu)^i & \text{if}\;\; 0.1 <\mu < 1, \\
e_0 + e_1 \mu^{-1/3} & \text{if}\;\; \mu < 0.1,
\end{cases}
\end{equation}
with the coefficients $d_i$ and $e_i$ given in Table \ref{tab_nu_app}.
Filled circles in Figure \ref{pic_nu_ups_mu}\textit{a} represent 
the results of \citet{sin83} under the double approximation, which 
are within $\sim10\%$ of our fully self-consistent values.
While \citet{sin83} found that $\upsilon_{c}$ monotonically increases 
with $\mu$ $(\simgt 0.09$),  our results show that 
$\upsilon_{c}$ attains a minimum at $\mumin\sim0.13-0.14$ for $n\simgt1.5$.  

For $\mu > \mumin$, the rotational effect is more important 
than the tidal force in maintaining equilibria of critical components 
(e.g., \citealt{rus28}).\footnote{\citet{rus28}
considered the ratio of the tidal to centrifugal forces at the 
\emph{undistorted} stellar surface, and found that tidal and rotational
effects are the same when $\mu=1/6$, which is not much different
from $\mumin$ that we have found.}
As $\mu$ increases, star \I\ would feel a weaker tidal force from its 
companion if the orbital distance were the same.  
Therefore, star \I\ must rotate faster to be critical, 
increasing $\upsilon_{c}$. This in turn corresponds to 
decreasing $\nu_c$ due to the condition of spin-orbit synchronization.
In the limit of $\mu\rightarrow1$,  $\upsilon_{c}$ converges to 
the value found in \S\ref{sec_rot} for critical rotation in the absence of 
the tidal force, while $\nu_c$ is tending to $\xi_{e}$, suggesting that 
a fictitious secondary with vanishing mass is located 
at the equator of the primary in critical condition.

\begin{figure*}[c!]
\epsscale{1.00}
\plotone{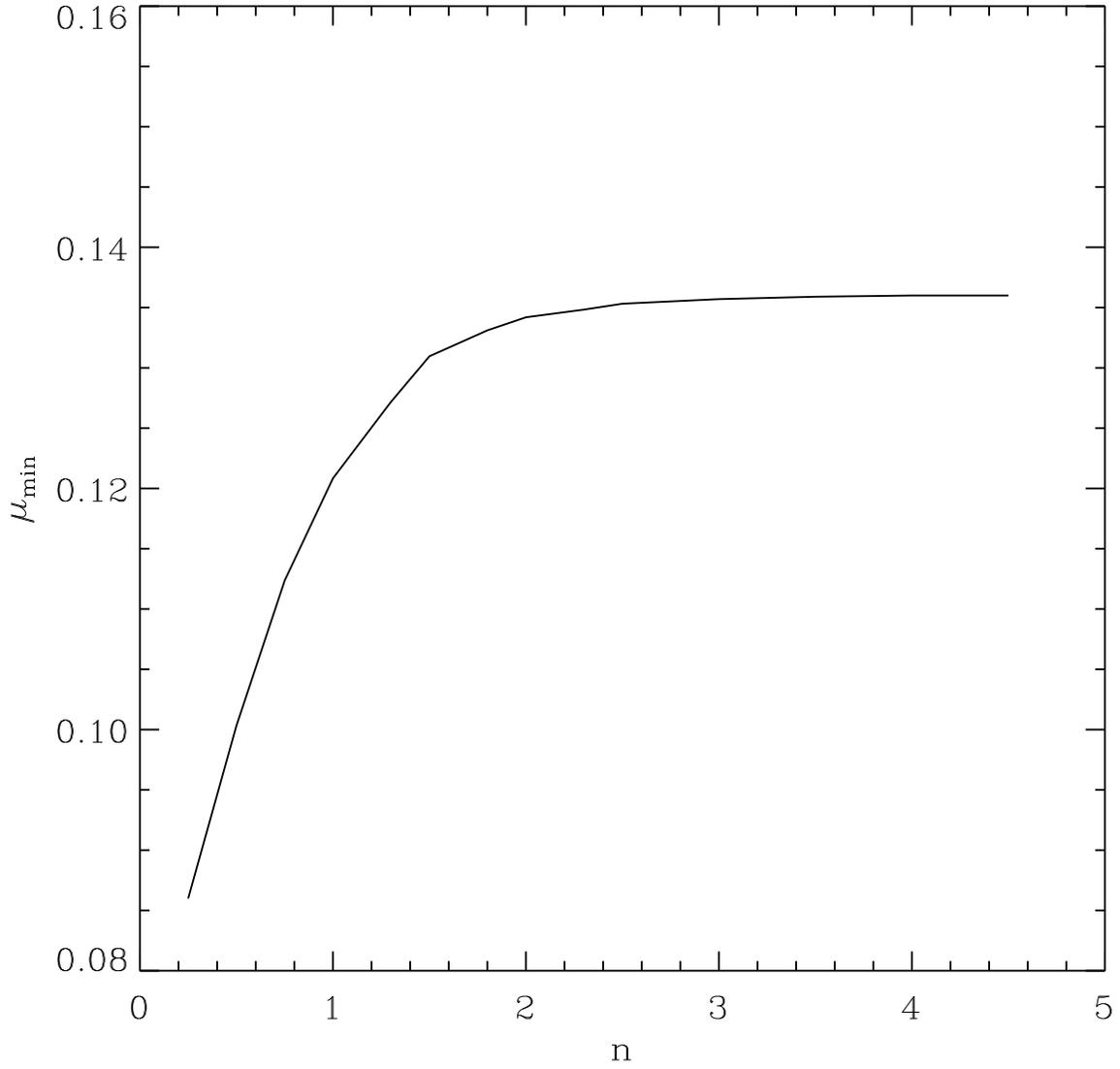}
\caption{Dependence on $n$ of $\mumin$ that minimize $\upsilon_c$.
Note that $\mumin\rightarrow 0.136$ as $n \rightarrow 5$.
}\label{pic_mu_min}
\end{figure*}

When $\mu < \mumin$, on the other hand, the tidal force dominates 
the centrifugal force in the force balance.  
Note that $\upsilon$ is proportional to
the ratio of the tidal force to the self-gravity at $r \approx R_1$,
so that $\upsilon_c \propto (\xi_1/\nu_c)^3/\mu$. 
Eggleton's formula (eg.\ [\ref{eq_eggleton}]) shows that $\xi_1/\nu_c$ 
decreases slightly more slowly than $\mu^{1/3}$ as $\mu$ decreases.  
This causes $\upsilon_c$ to increases slowly as 
$\mu$ decreases from $\mumin$.  In the limit of $\mu\rightarrow0$,
$\nu_c \propto \mu^{-1/3}$, and $\upsilon_c$ converges to 
$1.1530\times 10^{-2}$  and $1.2496\times 10^{-3}$ for the cases
with $n=1.5$ and $3.0$, respectively.\footnote{For the Roche model,
\citet{hor04} showed that the mean critical rotation
parameter is $\lambda_c \approx 2^4/3^5$ 
for a vanishingly small $\mu$ (see also \citealt{kop89}).
This corresponds to $\upsilon_c =\lambda_c/(\rho_c/\brho)
\approx1.002\times 10^{-2}$ and $1.041\times 10^{-3}$, 
since $\rho_c/\brho=6.5721$ and 63.2685 when $\mu\rightarrow0$ 
for $n=1.5$ and 3.0, respectively. These are in reasonable
agreement with our results for the critical components.}
Figure \ref{pic_mu_min} plots the variation of $\mumin$ with the
polytropic index, which shows that $\mumin = 0.136$
as $n$ approaches $5$.\footnote{Using the
Roche model, \citet{kop78} found the minimum value of $\lambda_c$ 
occurs in between $\mu=0.130$ and $0.167$ (or $q=0.15$ and 2), 
which is consistent with our results.}
For lower values of $n$, the enhanced equatorial radii due to 
flattening (e.g., Fig.\ \ref{pic_shape}\textit{d}) make the centrifugal
force more important at the surface, tending to decrease $\mumin$.

\subsubsection{Structure Constant $k_2$}\label{ap_k2}

Once the three-dimensional density distributions of disturbed polytropes 
are found, it is straightforward to calculate the apsidal motion 
constant $k_2$ defined by
\begin{equation}\label{eq_k2}
k_2=\frac{3-\eta_2(\xi_1)}{4+2\eta_2(\xi_1)},
\end{equation}
where $\eta_2(\xi_1)$ is the solution at the volume radius 
of the Radau equation that describes the distortion of 
equidensity surfaces inside a perturbed polytrope
(see, e.g., \citealt{kop78}).  For isolated polytropes,
we have confirmed that equation (\ref{eq_k2}) yields 
$\log\kiso=-0.8438$, $-1.8403$, and $-2.3081$ for $n=1.5$, 3.0, and 3.5 
cases, respectively, which are in good agreement with the published results
(e.g., \citealt{bro55}).

Figure \ref{pic_dlogk2} plots as solid lines the changes of 
$\Delta\log{k_2}\equiv \log(k_2/\kiso)$ with respect to the 
rotation parameter $\lambda$ and the central density concentration $\Delta 
\log \rho_c/\brho\equiv \log[(\rho_c/\brho)/(\rho_c/\brho)_{\rm iso}]$ 
for distorted polytropes with $n=1.5$, 3.0, and 3.5. 
Clearly, $k_2$ decreases as the perturbation amplitude
(or, the degree of central concentration) increases.  
Using realistic ZAMS models with modern opacities for unperturbed stars, 
but by relying on a quasi-spherical or double-approximation method 
in evaluating the sole effect of stellar rotation,
\citet{sto74} and \citet{cla99} found 
$\Delta\log{k_2}=-0.7\lambda$ and $\Delta\log{k_2}=-0.87\lambda$,
respectively. These are plotted as dotted and dashed lines in
the left panels of Figure \ref{pic_dlogk2}, 
approximately enveloping our results for $n=3.5$.  While it is not
straightforward to compare our results with those of 
\citet{sto74} and \citet{cla99} since the unperturbed
stellar models are different, this suggests that 
rotationally-distorted stellar models considered by these authors likely have 
internal structures similar to those of $n=3.5$ polytropes. 
The right panels of Figure \ref{pic_dlogk2} show that 
for polytropes with fixed $n$,
$\Delta\log{k_2}$ varies approximately linearly with 
$\Delta\log\rho_c/\brho$, nearly independent of $\mu$, with an 
average slope of $-1.12$, $-1.33$, and $-1.39$ for $n=1.5$, 
3.0, and 3.5 cases,
respectively.  This confirms the notion that 
$k_2$ is inversely proportional to the central density
concentration \citep{sto74}.
Figure \ref{pic_dlogk2} also shows that 
$|\Delta\log{k_2}|<0.1$ for $\mu\simlt0.9$, 
while it can be as large as 0.3 for strongly centrally-condensed 
polytropes in critical rotation with $\mu=1$.  
This suggests that the decrease in $k_2$ due to both rotational and tidal 
distortions is much smaller in binary stars with synchronous rotation
than in purely rotating stars.

\section{Apsidal Motion}\label{sec_apside}

Stars in close binaries have a shape that deviates 
from a spherical configuration, rendering the orbits of their companions 
non-Keplerian.
As mentioned in the Introduction, it is well known that tidal and
rotational distortions cause the apsidal line to advance, the rate of which 
depends on the degree of stellar concentration.   
Many observational studies have employed the analytic formulae of 
\citet{cow38} and \citet{ste39}, which are an eccentric extension 
of \citet{rus28} formula applicable to near-circular orbits.
By retaining only the linear-order terms, however, all of these
formulae are valid in a strict sense when the tidal and 
rotational distortions are very small.  Even in the cases when the effects 
of dynamical tides and relativity are unimportant, however, the distortion 
of a stellar shape can be substantial especially for critical components.
In this section, we check the validity of the classical formula
of \citet{rus28} using self-consistent solutions for 
stars in circular-orbit binaries for which tides are stationary.

\begin{figure*}
\epsscale{1.00}
\plotone{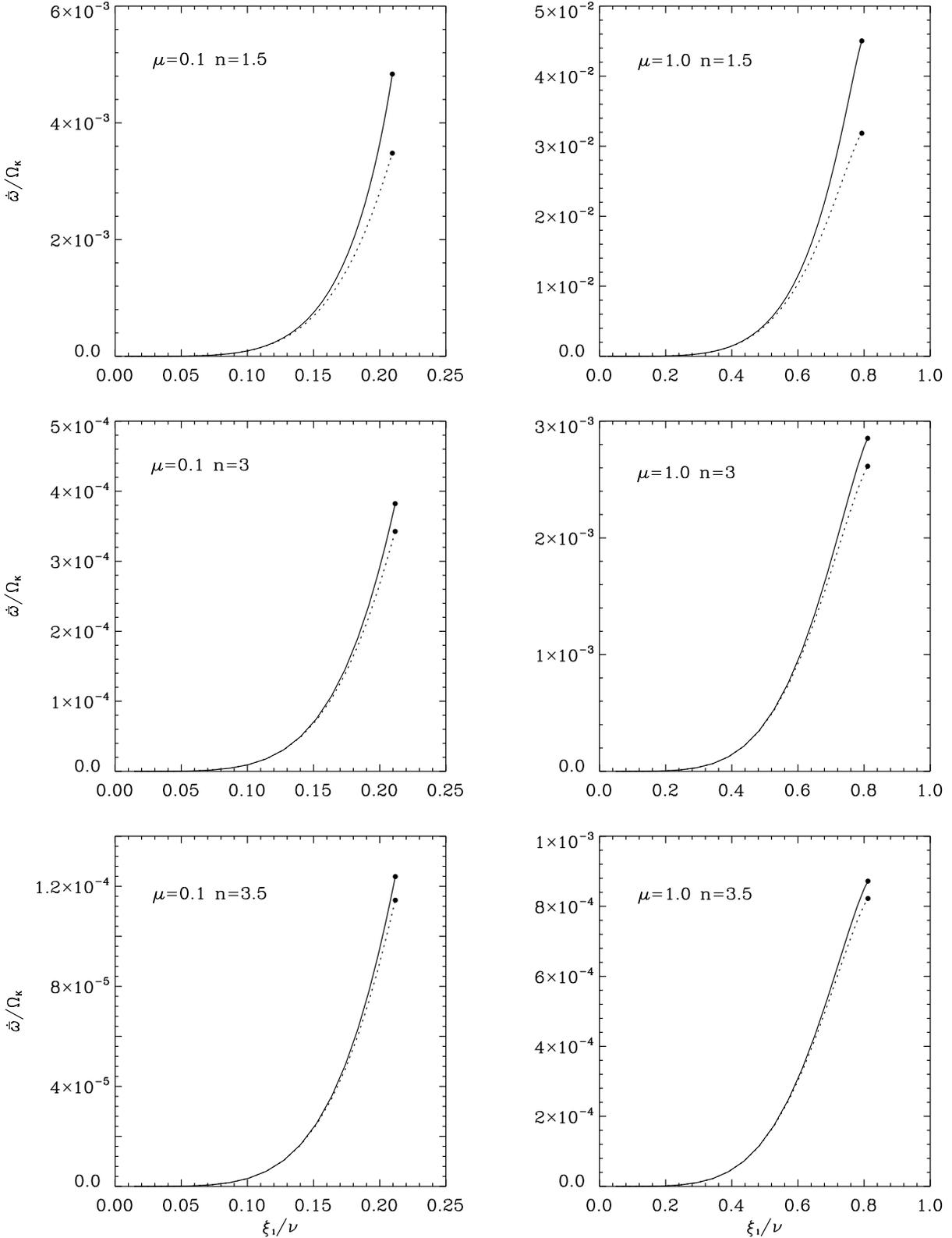}
\caption{Apsidal motion rate $\apside$ versus the relative radius $\xi_1/\nu$
for cases with (\textit{left}) $\mu=0.1$ and (\textit{right}) $\mu=1.0$. 
Solid lines plot $\apside_n$ calculated from our self-consistent 
solutions for distorted polytropes, while dotted lines are for
the theoretical rates $\apside_t$ based on the first-order perturbation 
method, with $k_2$ calculated for distorted polytropes from the Radau 
equation. For sufficiently large $\xi_1/\nu$, $\apside_n$ is larger 
than $\apside_t$.}
\label{pic_apse_ra_fmu} 
\end{figure*}

To obtain the apsidal motion rates from our distorted polytropes,
we calculate the self-gravitational potential $\Phi_g$ numerically along 
the axis joining the two binary components after the 
self-consistent solutions are obtained. 
We then apply the orbit theory in which the rate of change of the periastron 
is given by     
\begin{equation}\label{eq_aps_def}
\apside=\Omega-\kappa,
\end{equation}
where 
\begin{equation}\label{eq_kappa_def}
\kappa\equiv \frac{M}{M_1A^3} 
\frac{\partial(x^3\Phi_g)}{\partial x}\Biggr|_{\mathbf{x}=(A,0,0)}
\end{equation}
is the epicycle frequency
at the location of star \II\ (e.g., \citealt{mur99}).  
The resulting (numerical) apsidal motion rates, $\apside_n$,
versus the relative radius $R_1/A=\xi_1/\nu$ 
are shown as solid lines in Figure \ref{pic_apse_ra_fmu}
for $\mu=0.1$, $1.0$ and $n=1.5$, 3.0, 3.5 cases.
Also plotted as dotted lines are the (theoretical) rates, $\apside_t$, 
from the classical formula (eq.\ [\ref{eq_apsidal}]) 
with $k_2$ calculated in \S\ref{ap_k2}.
When the relative radius is small (i.e., $\xi_1/\nu < 0.15$ for $\mu=0.1$
and $\xi_1/\nu < 0.5$ for $\mu=1$) and thus the rotational and tidal 
perturbations are weak, the agreement between $\apside_n$
and $\apside_t$ are remarkably good.
As the amplitude of perturbations increases further, however, 
the classical formula begins to underestimate the real apsidal motion 
rates. This is because the the distortion of star \I\ becomes
no longer in the linear regime as $\xi_1/\nu$ approaches the
critical values.  In addition, we found that 
the gravitational potential due to a highly distorted
component has an non-negligible contribution from terms proportional 
to $x^{-2}$, while the classical theory took only $x^{-3}$ or higher-order 
terms (see, e.g., \citealt{ste39}), 
making $\apside_t$ smaller than $\apside_n$.

\begin{figure*}
\epsscale{1.00}
\plotone{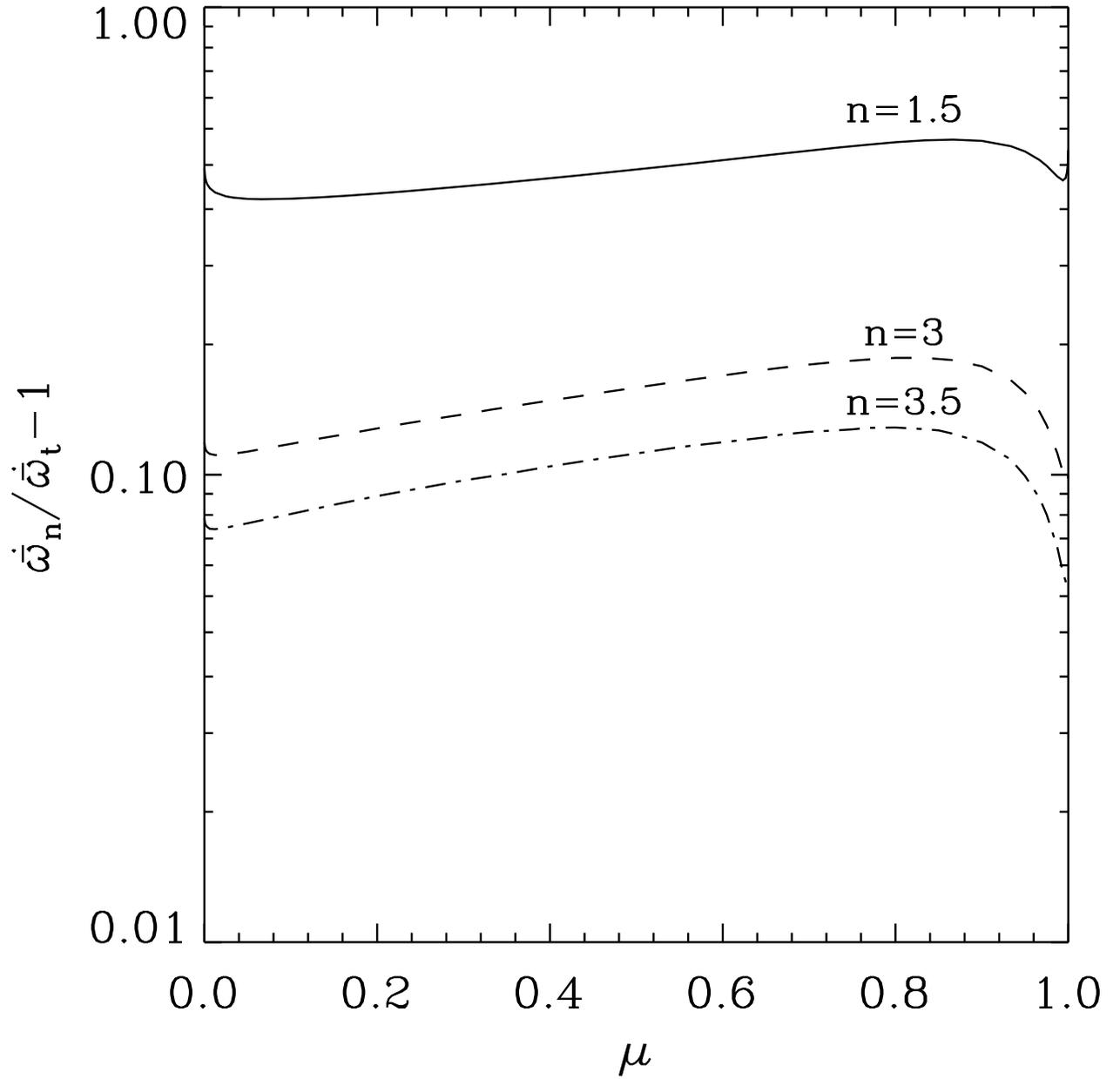}
\caption{Relative difference between the numerical and theoretical
apsidal motion rates, $\apside_n$ and $\apside_t$, 
against the relative mass $\mu$ for critical components. 
}\label{pic_apsec_mu}
\end{figure*}

Figure \ref{pic_apsec_mu} plots the relative difference 
$(\apside_n- \apside_t)/\apside_t$ of the numerical and theoretical rates of
the apsidal motions for critical configurations 
as a function of $\mu$.
For critically distorted polytropes with $n=1.5$, the real apsidal motion
rates are larger by $\sim 50\%$ than the values predicted by the 
classical theory.  As $n$ increases, the fractional mass occupying 
outer layers that exhibit large distortions decreases, yielding 
$(\apside_n- \apside_t)/\apside_t \sim 5-15\%$ for stars with $n=3.5$. 
These results suggest that the classical formula still works quite well 
for stars with large density concentration at centers, while it underestimates 
the apsidal motion rates significantly for stars with relatively low
$\rho_c/\brho$, such as in low-mass convective stars. 

\begin{figure*}
\epsscale{1.0}
\plotone{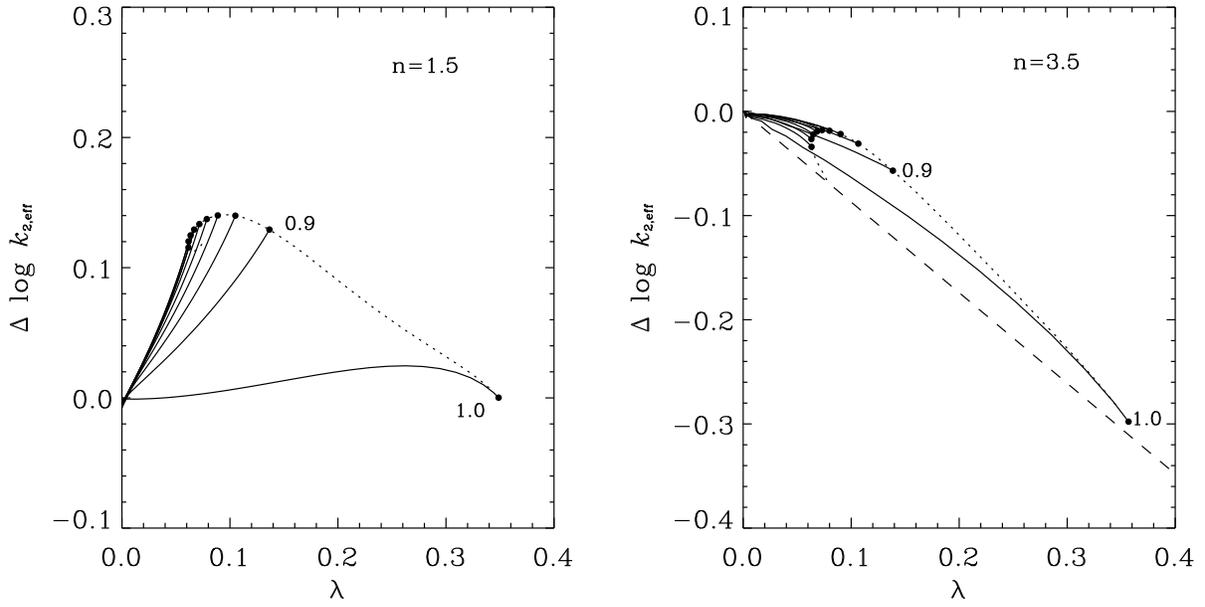}
\caption{ 
Dependence of $\Delta \log\keff$ on the rotation parameter $\lambda$
for (\textit{left}) $n=1.5$ and (\textit{right}) $n=3.5$ polytropes.
Each solid line is a sequence of distorted polytropes with
fixed relative mass $\mu$.  
Dotted lines correspond to the critical components.
The result, $\Delta \log\keff=-0.87\lambda$, of  \citet{cla99}
for rotationally-disturbed stars 
is drawn as a dashed line in the right panel.
}\label{pic_kcor}
\end{figure*}

Finally, we define the ``effective'' internal structure constant 
$\keff$ through
\begin{eqnarray} \label{eq_eff_aps_cor}      
\frac{\apside_n}{\Omega_K} = \keff
\Biggl(\frac{\xi_1}{\nu}\Biggr)^{5} 
\Biggl({1+7\frac{1-\mu} {\mu}} \Biggr).
\end{eqnarray}
We note that  $\keff$ introduced in equation (\ref{eq_eff_aps_cor})
is simply to match the real apsidal motion rates by means of the classical 
theory with the modified values of $k_2$, and thus it should 
not be interpreted as conveying physical information on the internal
structure of perturbed stars.  
Figure \ref{pic_kcor} shows the dependence of  
$\Delta\log\keff \equiv \log(\keff/k_{2,\rm iso})$ on $\lambda$ 
for $n=1.5$ and 3.5 polytropes. 
Dotted lines correspond to the cases with critically distorted polytropes.
Note that $\Delta\log\keff$ even increases with increasing $\lambda$ when
$n$ is small, although this does not mean stars become less centrally-condensed
as the amplitude of perturbations increases. 
For practical uses, we fit $\keff$ using 
\begin{equation} \label{eq_kcor_appr}
\Delta\log\keff=a_1\lambda+a_2\lambda^2+a_3\lambda^3,
\end{equation}
with the coefficients $a_i$ listed in Table \ref{tab_k2eff} 
for $n=1.5$, 3.0, and 3.5 cases.
The relative errors of the fits are less than 2\%.

\section{Summary and Discussion}

Binary stars are a traditional source of information on stellar mass,
radius, luminosity, etc., the inter-relationships of which are of crucial 
importance to stellar astrophysics.   Many observational and theoretical
arguments suggest that centrifugal and tidal forces may 
significantly affect the physical properties of stars in close binary
systems.  Since the pioneering work of \citet{cha33a,cha33b,cha33c},
there have been numerous theoretical studies that tried to find 
equilibrium solutions of polytropic stars under rotational and tidal 
perturbations.  However, most of them take some type of approximations, 
and thus are valid only when the perturbations are weak.
In order to construct equilibrium structures of such polytropes without 
relying on 
any assumption on the perturbation strength and the degree of stellar 
distortion, we use an SCF method in which the angular orbital velocity 
$\Omega$
is self-consistently constrained rather than being treated as a 
free parameter.  

To ignore the complicated effects of time-dependent tides and to make 
contact with the previous theoretical work, we consider a circular-orbit, 
synchronized binary consisting of a polytrope and a point-mass object. 
Based on our SCF scheme described in \S\ref{sec_for},
we solve the dimensionless steady-state momentum equation 
(\ref{eq_balance_dim}) and Poisson equation (\ref{eq_Pos_dless}) 
in the rotating frame at $\Omega$, alternately and iteratively, 
until the converged solutions are obtained.
We first apply our method to polytropes in isolation and  
subsequently in uniform rotation,
and calculate the density concentration $\rho_c/\brho$,
the volumn radius $\xi_1$, the equatorial radius $\xi_e$, and the polar radius 
$\xi_p$; these are in good agreement with  the accurate results
of \citet{cha39} for unperturbed polytropes and \citet{jam64} for 
rotating polytropes, confirming the performance of our SCF method.  
When the rotation parameter $\upsilon$ (eq.\ [\ref{eq_ups_mu_nu}])
is sufficiently large, a rotating polytrope reaches a critical state,
beyond which no equilibrium is possible.
Equation (\ref{eq_rot_fit}) and Table \ref{tab_rot_fit} provide fitting
formulae of $\xi_1$, $\xi_e$, $\xi_p$, $\rho_c/\brho$, $N_n$, and 
the moment of inertia $I$ for polytropes with $n=1.5$ or 3. 
By comparing the axis ratio 
$\mathcal{R}=\xi_e/\xi_p$ and the velocity ratio $\vel_e/\vel_c$
with the predictions from the conventional Roche model, we find that 
in terms of the critical value $\mathcal{R}_c$, the Roche model is 
a reasonable approximation for rapidly rotating stars with high central 
density concentration (i.e., when $n\simgt 2.0$), although it underestimates 
$\vel_e/\vel_c$ by about 10\% when $n\simgt 3.0$.

We then calculate equilibrium solutions of polytropes subject to both 
centrifugal and tidal forces.  The rotational and tidal perturbations 
in general make
disturbed stars larger in size and more condensed toward their centers.
For $n=1.5$ polytropes in critical configurations, for example, 
the distorted stars are 
larger by $\sim23\%$ and $\sim4-8\%$ for $\mu=1$ and $\mu\sim0.1-0.9$,
respectively, than the unperturbed counterparts with the same specific 
entropy.  Here, $\mu$ is the mass of a perturbed star relative to 
the total mass of a binary system.  
For more centrally-concentrated polytropes with $n=3.5$, the enlargement 
factor is reduced to 
$\sim12\%$ and $\sim2-3\%$ for $\mu=1$ and $\mu\sim0.1-0.9$,
respectively.  
In determining the structure of a critical component,
the rotational effect is more important than the tidal effect 
provided $\mu>\mumin\sim0.13-0.14$. 
As in purely rotationally-distorted cases, the shape and size of 
both rotationally- and tidally-perturbed polytropes are well
described by the Roche models as long as $n\simgt 2$.
Equation (\ref{eq_ra_fit}) and Tables \ref{tab_rho_app},
\ref{tab_xi_app}, \ref{tab_Nn_app}, and \ref{tab_I_app} give
functional fits of $\rho_c/\brho$, $\xi_1$, $N_n$, and $I$ 
on the mass radio $\mu$ and the relative radius $R_1/A=\xi_1/\nu$, while
equation (\ref{eq_jj_fit}) is for the fit of $\Jspin/\Jorb$,
the ratio of spin to orbital angular momenta of star \I.
Equations (\ref{eq_rhoc_fit})-(\ref{eq_crit_nu}) 
and Table (\ref{tab_nu_app}) give the dependences 
on $\mu$ of $\rho_c/\brho$, 
$N_n$, $\Jspin/\Jorb$, $I$, $\upsilon_c$, and $\nu_c$ for 
critical configurations. 

Using self-consistent equilibrium solutions of rotationally- and 
tidally-distorted polytropes, we calculate the internal structure constant 
$k_2$ by integrating the Radau equation as well as the apsidal motion rate
$\apside_n$ based on the orbit theory.  We find that 
$\Delta\log k_2$ depends almost linearly upon 
$\Delta\log\rho_c/\brho$, with an average slope of
$-1.12$, $-1.33$, and $-1.39$, nearly independent of $\mu$,  
for $n=1.5$, 3.0, and 3.5 polytropes, respectively.
The expression $\Delta\log k_2\sim-(0.7-0.87)\lambda$ 
found for rotating stars from
\citet{sto74} and \citet{cla99} appears to be valid only for
stars with very high density concentration, corresponding to $n\approx 3.5$.
The comparison of $\apside_n$ calculated from 
our SCF method with $\apside_t$ from the classical formula of
\citet{rus28} based on the first-order perturbation analyses 
shows that the latter is valid
only when the amplitude of perturbations is small. 
For the cases of critical configurations where tidal and centrifugal forces
are rather strong, the classical theory underestimates the real apsidal 
motions rates by as much as $\sim50\%$ for $n=1.5$ polytropes and
$\sim5-15\%$ for $n=3.5$, indicating that the nonlinear effects 
are by no means negligible in augmenting the apsidal motion rates. 
To this end, we define the effective internal structure constant $\keff$
that gives the real apsidal motion rates by means of the classical formula. 
Equation (\ref{eq_kcor_appr}) and Table \ref{tab_k2eff} present the
algebraic fits to $\keff$.

Observations of eclipsing binary stars indicate that B-type binary stars are 
about 20\% smaller in size than the single stars of the same spectral type, 
while A- and F-type binary stars are larger than the corresponding isolated 
stars (e.g., \citealt{mal03,mal07}).  
For M-type stars with mass $0.4-0.8$ M$_\odot$, 
\citet{rib06} reported that observed radii of eclipsing binary stars 
are larger by $5-10\%$ than those of the theoretical, isolated models. 
On the other hand, our self-consistent calculations show that centrifugal 
and tidal perturbations make binary stars bigger in size.  
These observational and theoretical 
results are qualitatively reconciled with each other if one considers the 
fact that early-type (late-type) binary stars rotate 
slower (faster) than the isolated counterparts
(e.g., \citealt{mcn65,mal93,mal03}).  Slow rotations of early-type stars
in close binaries are apparently a consequence of the spin-orbit synchronization
occurring during their main-sequence phase, induced by tidal torque due 
to either radiative damping of dynamical tides \citep{zah77} or large-scale 
mechanical currents \citep{tas87,tas88}.  
In addition, early-type binary stars appear to have a low
initial rotational velocity amounting to $\sim130$ km s$^{-1}$, 
compared to $\sim200$ km s$^{-1}$ for single stars, 
when they enter the ZAMS stage \citep{kha07}.

\begin{figure}
\epsscale{1.}
\plotone{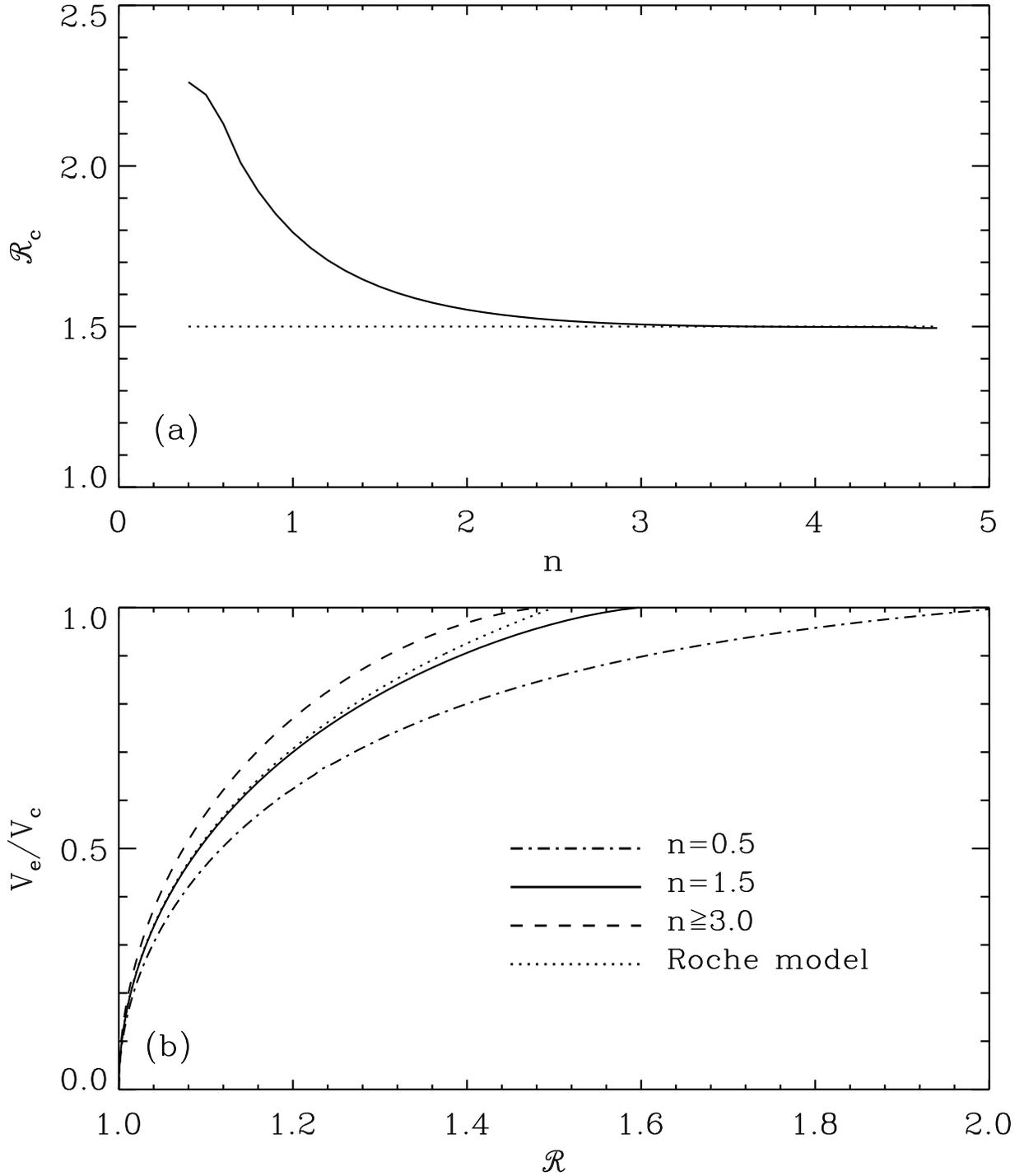}
\caption{\textit{a}: Critical values of the aspect ratio 
$\mathcal{R} =\xi_e/\xi_p$ resulting from our self-consistent solutions
for rotating polytropes with index $n$ (\textit{solid line})
and the prediction, $\mathcal{R}_c=1.5$, of the Roche models 
(\textit{dotted line}).
\textit{b}: Dependence on $\mathcal{R}$ of the ratio 
$V_e/V_c$ of the equatorial to critical velocities.  
Curves for $n\geq3.0$ are almost indistinguishable.
} \label{pic_crit}
\end{figure}

By analyzing the near-IR, long-baseline 
interferometric data of the Be star Achernar which is 
known as a rapid rotator, \citet{dom03} reported that the star has a 
projected aspect ratio of $\mathcal{R}=1.56\pm0.05$. 
Since the projection always tends to decrease the measured aspect ratio,
this implied that the real oblateness must be larger 
than the critical value under the Roche approximation. 
If Achernar can be modeled by a single, rigidly-rotating polytrope,
Figure \ref{pic_crit}\textit{a} suggests that it must have $n< 2.0$, which
is unlikely given that Be stars may have a larger central density 
concentration than that of a polytrope with $n=3$.
\citet{jac04} showed that differentially-rotating stellar models with a
more realistic equation of state allow equilibrium stars with 
$\mathcal{R}$ larger than 1.5, compatible with the interferometric
observations.
On the other hand, \citet{car08} found that the data are also
consistent with a critically-rotating central star surrounded by
a small equatorial disk produced by the ejected material from 
the central star.

If one component (star \I) in a semidetached binary overfills its Roche lobe,
it may rapidly lose mass to its companion in a few orbital periods.
Such dynamical mass transfer is likely to create a common envelope binary,
important for the evolution of cataclysmic binaries and double degenerate 
white dwarfs (e.g., \citealt{pac76,fra02}).  
In finding the condition for the onset of the dynamical mass transfer, 
most previous work considered the conservation of the orbital angular momentum,
neglecting the contribution of the spin angular momentum.
Using the Eggleton formula (eq.\ [\ref{eq_eggleton}]) for the Roche lobe
radius, for instance, \citet{hje87} found that the rapid conservative 
mass transfer from an $n=1.5$ polytrope to its compact companion occurs
if $q=M_1/M_2 > q_c=0.634$, while the use of the 
less accurate \citet{pac71} formula, $R_L/A=0.4622(q/[1+q])^{1/3}$, 
yields $q_c = 2/3$ \citep{rap82,dso06,mot07}.
We revisit this issue using the conservation of the total angular momentum 
that includes $\Jspin$ given by equation (\ref{eq_jj_fit}),
and find that the critical value of $q$ for the dynamical mass transfer
is increased to $q_c=0.693$ and 0.736 under the Eggleton and 
Paczy\'nski formulae, respectively, for $n=1.5$ polytropes.
Note that the value of $q_c$ based on the Eggleton formula including 
$\Jspin$ is similar to that under the Paczy\'nski formula without
the effect of $\Jspin$.  Although the changes in $q_c$ due to
the inclusion of the spin angular momentum are not
great since $\Jspin/\Jorb \simlt 0.1$ for $q<1$ (or $\mu<0.5$), the
effect of $\Jspin$ can be significant in  
the processes of mass transfer if the donor-to-accretor mass ratio is large.

\begin{figure*}
\epsscale{1.00}
\plotone{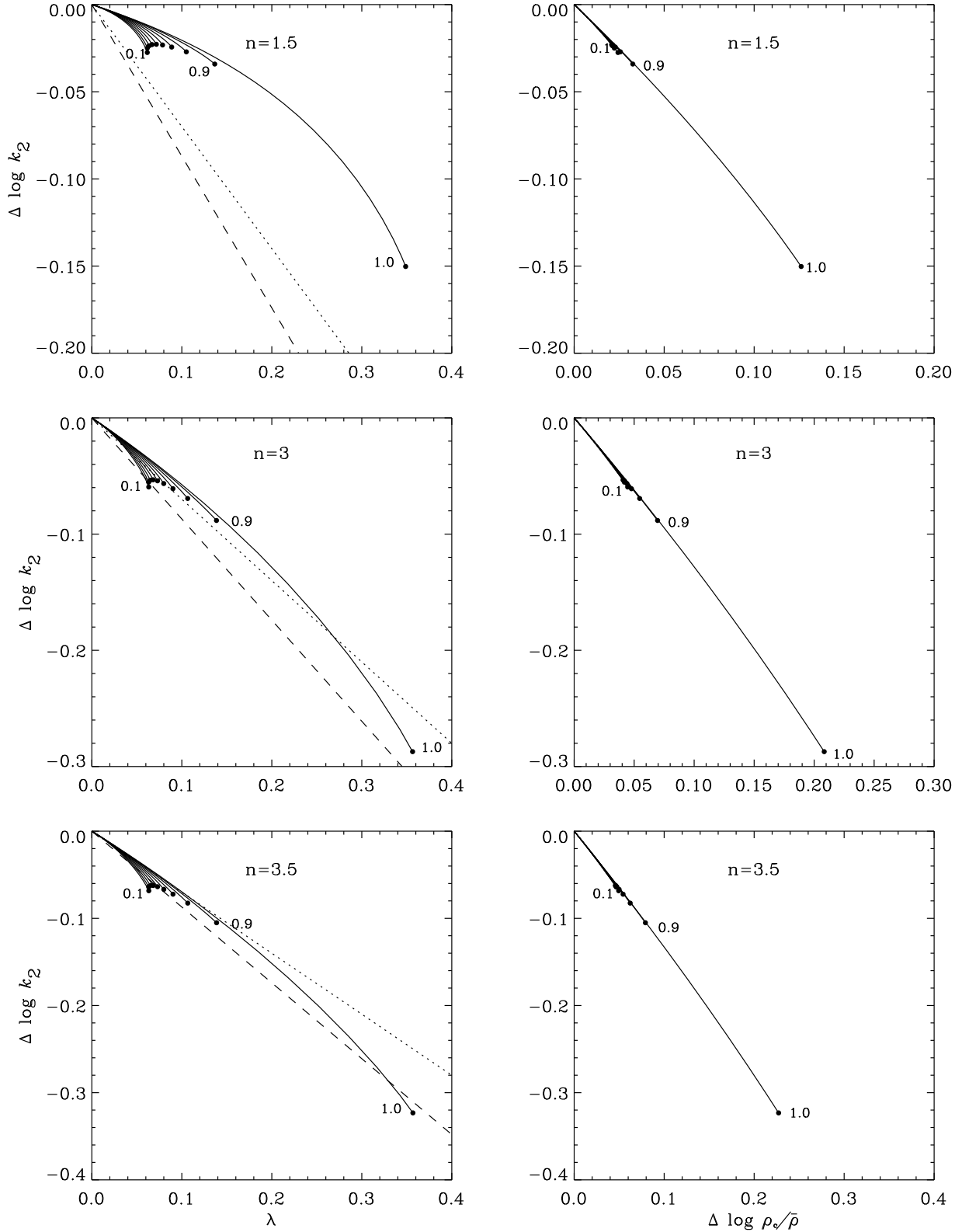}
\caption{Variations of $\Delta \log {k_2}$ with 
(\textit{left}) $\lambda$ and (\textit{right}) $\Delta \log \brho/\rho_c$
for polytropes with $n=1.5$, 3.0, and 3.5.  Each solid line is a sequence of 
distorted polytropes with fixed relative mass $\mu$, with a filled circle
at the tip of the line corresponding to a critical component.
For comparison, $\Delta \log {k_2}=-0.7\lambda$ 
and $-0.87\lambda$, the results of \citet{sto74} and \citet{cla99} 
for rotationally-disturbed stars, are 
plotted as dotted and dashed lines, respectively.
}\label{pic_dlogk2}
\end{figure*}

Measuring apsidal motion rates is probably a unique means to probe
the internal density concentration of binary stars.  For 
systems where the effects of general relativity \citep{gim85} or
a third body (e.g., \citealt{boz07}) are unimportant, most previous
studies adopted the classical theory of \citet{ste39} with 
$k_2$ calculated from isolated polytropes
and found that binary stars are much more centrally condensed than predicted
(e.g., \citealt{sch58,kop78}).  More recently, \citet{cla93} 
and \citet{cla03} showed the theoretical results are in good agreement
with observations if the changes of $k_2$ due to rotation as well as 
the effects of the initial chemical composition and convective core 
overshooting are taken into account.  However, the reduction of $k_2$ 
considered in these papers amounts to $\Delta\log{k_2}\sim -(0.2- 0.3)$,
which may be possible only for purely rotating cases with $\mu=1$. 
Even for highly centrally-concentrated polytropes with $n=3.5$,
Figure \ref{pic_dlogk2} shows that $|\Delta\log{k_2}| < 0.1$ for
synchronous binaries in the mass range of $0.1 \simlt\mu \simlt0.9$.  
In addition, our SCF analyses show that the classical formula derived
under the linear approximation is likely to underestimate the real 
apsidal motion rates for stars with low central concentration, 
especially in critical configurations.
All of these suggest that special care should be taken in interpreting 
observed data for apsidal motions of close binary systems where stellar
rotation is slow.

\acknowledgements
The authors acknowledge a helpful report from an anonymous referee, 
and Jeremiah P.\ Ostriker for a careful reading of the 
manuscript. 
This work was supported by the Creative Research Initiatives program, 
CEOU of MEST/KOSEF.

\clearpage

\begin{deluxetable}{ccccc}
  \tabletypesize{\scriptsize}
  \tablecaption {Parameters of undisturbed polytropes}
  \tablewidth{0pt}
  \tablehead {
   $n$   & $\xi_1$ & $\rho_c/\brho$&   $N_n$ &      $I$   }
  \startdata
   0.5 & 2.7527 &    1.8352    & 2.5235 & 117.575  \\
   1.0 & 3.1416 &    3.2899    & 0.6366 & 101.844  \\
   1.5 & 3.6538 &    5.9907    & 0.4242 &  93.156  \\
   2.0 & 4.3529 &   11.4025    & 0.3648 &  88.895  \\
   2.5 & 5.3553 &   23.4065    & 0.3515 &  88.190  \\
   3.0 & 6.8969 &   54.1825    & 0.3639 &  90.910  \\
   3.5 & 9.5358 &  152.8837    & 0.4010 &  98.398  \\
   4.0 & 14.972 &  622.4079    & 0.4772 & 114.275  \\
   4.5 & 31.837 & 6189.4719    & 0.6580 & 152.617 
  \enddata
\tablecomments{$n$: polytropic index;
$\xi_1$: dimensionless volume radius;
$\rho_c/\brho$: central density concentration;
$N_n$: coefficient of the mass-radius relation;
$I$: dimensionless moment of inertia}
\label{ta_unperturbed_comparison}
\end{deluxetable}

\begin{deluxetable} {cccccrcccc}
\tabletypesize{\scriptsize}
\tablecaption {Parameters of uniformly-rotating polytropes in 
critical configuration}
  \tablewidth{0pt}
  \tablehead {

    $n$   &           $\upsilon_c$ & $\lambda_c$ &  $N_n$ &      $I$  &
$\rho_c/\brho$&  $\xi_1$ &   $\xi_p$ & $\xi_e$ 
             }
  \startdata
    0.5 & 1.4409$\times10^{-1}$ & 0.29386  & 1.42922 & 587.284 &
     2.0394 &  3.4754 &   2.1779 &  4.8489 \\
    1.0 & 8.3880$\times10^{-2}$ & 0.33967  & 0.45025 & 236.366 &
     4.0494 &  3.7367 &   2.6922 &  4.8278 \\
    1.5 & 4.3622$\times10^{-2}$ & 0.35168  & 0.34330 & 153.088 &
     8.0619 &  4.2676 &   3.2962 &  5.3526 \\
    2.0 & 2.1576$\times10^{-2}$ & 0.35694  & 0.32178 & 120.783 &
    16.5434 &  5.0863 &   4.0556 &  6.2961 \\
    2.5 & 9.9305$\times10^{-3}$ & 0.35922  & 0.32965 & 106.637 &
    36.1737 &  6.3014 &   5.1000 &  7.7549 \\
    3.0 & 4.0803$\times10^{-3}$ & 0.36019  & 0.35728 & 102.111 &
    88.2755 &  8.1907 &   6.6738 & 10.0548 \\
    3.5 & 1.3861$\times10^{-3}$ & 0.36057  & 0.40779 & 105.249 &
   260.1278 & 11.4334 &   9.3419 & 14.0214 \\
    4.0 & 3,2913$\times10^{-4}$ & 0.36071  & 0.49864 & 118.369 &
  1095.9536 & 18.1078 &  14.8100 & 22.1991 \\
    4.5 & 3.2295$\times10^{-5}$ & 0.36074  & 0.70213 & 154.862 &
 11170.2095 & 38.7737 &  31.7202 & 47.5299 \\
  \enddata
\tablecomments{
$n$: polytropic index;
$\upsilon_c$ \& $\lambda_c$ : critical 
values of the rotation parameters;  
$N_n$: coefficient of the mass-radius relation;
$I$: dimensionless moment of inertia;
$\rho_c/\brho$: central density concentration;
$\xi_1$, $\xi_p$, \& $\xi_e$:
dimensionless volume, polar, and equatorial radii, respectively
}\label{ta_rot_terminated_values}
\end{deluxetable}

\begin{deluxetable} {crrrrrcrrrrr}
\tabletypesize{\scriptsize}
\tablecaption {Fitting coefficients for the various parameters 
of rotating polytropes}
\tablewidth{0pt}
\tablehead {
   & \multicolumn{5}{c}{$n=1.5$} && \multicolumn{5}{c}{$n=3$}   \\
            \cline{2-6} \cline{8-12}\\
$f$  
  &   $a_0$ &      $a_1$ &      $a_2$ &       $a_3$ &    $a_4$  & 
  &   $a_0$ &      $a_1$ &      $a_2$ &       $a_3$ &    $a_4$ 
             }
  \startdata
 $\xi_1$
               & 3.65375 &  0.03588 &  0.00452 & $-0.11724$  & 0.21911 &
               & 6.89685 &  0.30538 &  0.18974 & $-0.24088$  & 0.96237 \\
 $\xi_e$
               & 3.65375 &  0.06834 &  0.01192 & $-0.25324$  & 0.22619 &
               & 6.89685 &  0.40291 &  0.47814 & $-0.45359$  & 0.99175 \\
 $\xi_p$
               & 3.65375 &$-0.07555$&$-0.00142$&        0    & 0       &
               & 6.89685 &$-0.22243$&$-0.00057$&        0    & 0       \\
$\rho_c/\brho$
               & 5.99070 &  0.01990 &  0.01092 & $-0.52212$  & 0.22119 &
               &54.18248 &  3.83480 &  4.55374 & $-7.62154$  & 0.96447 \\
  $N_n$
               & 0.42422 &$-0.00904$&$-0.00057$& 0.00996     & 0.21821 &
               & 0.36394 &$-0.00599$&$-0.00066$&        0    & 0       \\
   $I$
               & 93.15635&  2.05830 &  0.25902 &$-30.87279$  & 0.17755 &
               & 90.90975&  8.53385 &  2.62075 &        0    &       0
  \enddata
\tablecomments{$a_i$'s are defined through equation (\ref{eq_rot_fit}).}
\label{tab_rot_fit}
\end{deluxetable}

\begin{deluxetable} {crrrrcrrrr}
\tabletypesize{\scriptsize}
\tablewidth{0pt}
\tablecaption{
Fitting coefficients for $\rho_c/\brho$ of
distorted polytropes}
\tablehead {
 &\multicolumn{4}{c} {$n=1.5$, $b_0=5.99070$} &
 &\multicolumn{4}{c} {$n=3$,   $b_0=54.18248$} \\
             \cline{2-5} \cline{7-10}\\
 $\mu$&      $b_1$ &   $b_2$ &  $b_3$ &    $b_4$&  
      &      $b_1$ &   $b_2$ &  $b_3$ &    $b_4$ }
 \startdata
   0.1&  $-0.17092$& 3.49535&  0.06614&  28.20734&
      &  $-2.46052$& 59.12506& 9.44135&  17.55872\\
   0.2&  $-0.12860$& 2.13006&  0.03425&  21.79792&
      &  $-1.85425$& 36.09363& 5.38216&  13.12862\\
   0.3&  $-0.10830$& 1.58568&  0.02566&  18.13797&
      &  $-1.55648$& 26.79729& 4.29253&  10.56543\\
   0.4&  $-0.09552$& 1.28070&  0.02340&  15.36396&
      &  $-1.36230$& 21.48995& 4.03468&  ~8.63850\\
   0.5&  $-0.08637$& 1.08067&  0.02478&  12.96732&
      &  $-1.21555$& 17.90157& 4.23624&  ~7.02247\\
   0.6&  $-0.07921$& 0.93547&  0.03011&  10.74527&
      &  $-1.09285$& 15.18838& 4.84080&  ~5.60038\\
   0.7&  $-0.07320$& 0.82138&  0.04173&  ~8.62004&
      &  $-0.98314$& 12.95893& 5.89126&  ~4.33881\\
   0.8&  $-0.06148$& 0.68295&  0.08494&  ~6.11931&
      &  $-0.88629$& 11.05241& 7.38448&  ~3.25984\\
   0.9&  $-0.06648$& 0.65962&  0.09659&  ~4.98583&
      &  $-0.83960$& ~9.68781& 8.75499&  ~2.46508
 \enddata
\tablecomments{$b_i$'s are defined through equation (\ref{eq_ra_fit}).}
\label{tab_rho_app}
\end{deluxetable}

\begin{deluxetable} {crrrrcrrrr}
\tabletypesize{\scriptsize}
\tablewidth{0pt}
\tablecaption{Fitting coefficients for the volume radius $\xi_1$ of
distorted polytropes}
\tablehead {
  &\multicolumn{4} {c} {$n=1.5$, $b_0=3.65375$} &
  &\multicolumn{4} {c} {$n=3$,  $b_0=6.89685$}  \\ 
             \cline{2-5} \cline{7-10}\\
 $\mu$&      $b_1$ &   $b_2$ &  $b_3$ &    $b_4$&  
      &      $b_1$ &   $b_2$ &  $b_3$ &    $b_4$   
}
 \startdata
    0.1&  $-0.05174$& 1.22025&  0.13999& 19.60622 &
       &  $-0.10449$& 2.60661&  0.64437&  15.45073\\
    0.2&  $-0.03954$& 0.75031&  0.07579& 14.87539 &
       &  $-0.07951$& 1.59668&  0.36468&  11.50900\\
    0.3&  $-0.03346$& 0.56039&  0.05810& 12.14228 &
       &  $-0.06669$& 1.18356&  0.29108&   9.18447\\
    0.4&  $-0.02956$& 0.45272&  0.05287& 10.08102 &
       &  $-0.05814$& 0.94518&  0.27414&   7.41921\\
    0.5&  $-0.02668$& 0.38076&  0.05430&  8.32835 &
       &  $-0.05149$& 0.78149&  0.28845&   5.93073\\
    0.6&  $-0.02434$& 0.32725&  0.06149&  6.75355 &
       &  $-0.04571$& 0.65486&  0.33023&   4.61883\\
    0.7&  $-0.02233$& 0.28415&  0.07542&  5.31595 &
       &  $-0.04026$& 0.54723&  0.40290&   3.45588\\
    0.8&  $-0.01862$& 0.23208&  0.11519&  3.75473 &
       &  $-0.03496$& 0.44995&  0.50912&   2.45910\\
    0.9&  $-0.02015$& 0.22374&  0.11821&  3.07080 &
       &  $-0.03043$& 0.36682&  0.62743&   1.69514
\enddata
\tablecomments{$b_i$'s are defined through equation (\ref{eq_ra_fit}).}
\label{tab_xi_app}
\end{deluxetable}

\begin{deluxetable} {crrrrcrrrr}
\tabletypesize{\scriptsize}
\tablewidth{0pt}
\tablecaption{
Fitting coefficients for $N_n$ of
distorted polytropes}
\tablehead {
 &\multicolumn{4} {c} {$n=1.5$, $b_0=0.42422$} &
 &\multicolumn{4} {c} {$n=3$, $b_0=0.36394$} \\
             \cline{2-5} \cline{7-10}\\
 $\mu$&    $b_1$ &   $b_2$ &  $b_3$ &  $b_4$&  
      &    $b_1$ &   $b_2$ &  $b_3$ &  $b_4$      }
 \startdata
  0.1&  0.00696& $-$0.18218& $-$0.06256&   14.81447&
     &  0.00110& $-$0.06453&    1.14963& $-$11.8357\\
  0.2&  0.00533& $-$0.11239& $-$0.03379&   11.12954&
     &  0.00086& $-$0.03997&    0.56067& $-$9.41496\\
  0.3&  0.00452& $-$0.08394& $-$0.02569&    8.96846&
     &  0.00075& $-$0.03009&    0.36519& $-$8.22768\\
  0.4&  0.00399& $-$0.06755& $-$0.02301&    7.32661&
     &  0.00068& $-$0.02458&    0.26786& $-$7.45925\\
  0.5&  0.00357& $-$0.05633& $-$0.02307&    5.92820&
     &  0.00064& $-$0.02104&    0.20972& $-$6.89082\\
  0.6&  0.00321& $-$0.04771& $-$0.02524&    4.67990&
     &  0.00061& $-$0.01856&    0.17112& $-$6.43430\\
  0.7&  0.00288& $-$0.04047& $-$0.02957&    3.55301&
     &  0.00059& $-$0.01675&    0.14367& $-$6.04829\\
  0.8&  0.00229& $-$0.03146& $-$0.04086&    2.37704&
     &  0.00058& $-$0.01534&    0.12333& $-$5.72063\\
  0.9&  0.00238& $-$0.02892& $-$0.04344&    1.78659&
     &  0.00055& $-$0.01410&    0.10903& $-$5.50715
 \enddata
\tablecomments{$b_i$'s are defined through equation (\ref{eq_ra_fit}).}
\label{tab_Nn_app}
\end{deluxetable}

\begin{deluxetable} {crrrrcrrrr}
\tabletypesize{\scriptsize} \tablewidth{0pt} \tablecaption{
Fitting coefficients for $I$ of distorted polytropes} \tablehead {
 &\multicolumn{4} {c} {$n=1.5$, $b_0=90.91016$} &
 &\multicolumn{4} {c} {$n=3$, $b_0=93.15765$} \\
             \cline{2-5} \cline{7-10}\\
 $\mu$&    $b_1$ &   $b_2$ &  $b_3$ &  $b_4$&
      &    $b_1$ &   $b_2$ &  $b_3$ &  $b_4$      }
 \startdata
0.1&    0.23389&    -6.78106&   416.21309&   0.06683&&   -4.44876&
133.13261&  152.14256&  10.18132\\
0.2&    0.13995&    -3.24888&   197.05415&   0.05320&&   -3.27895&
78.57307&   81.88516&   7.48309\\
0.3&    0.10835&    -2.25890&   127.38938&   0.00015&&   -2.69940&
56.50289&   58.48508&   6.01238\\
0.4&    0.09711&    -1.90272&   93.74429&   -0.06256&&   -2.32438&
43.95814&   47.13868&   4.97229\\
0.5&    0.09662&    -1.80771&   74.20831&   -0.12872&&   -2.05058&
35.62331&   40.77503&   4.14597\\
0.6&    0.10447&    -1.86636&   61.68649&   -0.19697&&   -1.83950&
29.58690&   36.91530&   3.45368\\
0.7&    0.12211&    -2.06525&   53.29763&   -0.26863&&   -1.68084&
25.05345&   34.30903&   2.86731\\
0.8&    0.15798&    -2.48941&   47.93900&   -0.35028&&   -1.48905&
20.90588&   33.51635&   2.31784\\
0.9&    0.25324&    -3.60727&   46.46436&   -0.47006&&   -1.67258&
20.43106&   27.78674&   2.09483\\
 \enddata
\tablecomments{$b_i$'s are defined through equation
(\ref{eq_ra_fit})} \label{tab_I_app}
\end{deluxetable}

\begin{deluxetable} {crr}
\tabletypesize{\scriptsize}
\tablewidth{0pt}
\tablecaption{
Fitting coefficients for $\nu_c$ of
the critical components}
\tablehead {
     &  $n=1.5$ &  $n=3$ }
\startdata
  $d_0$ &    5.283    &    10.051   \\
  $d_1$ &   22.07    &    39.75     \\
  $d_2$ & $-87.99$    & $-155.79 $  \\
  $d_3$ &  236.69     &   420.65    \\
  $d_4$ &$-292.31$    & $-522.14 $  \\
  $d_5$ &  141.03     &   253.75    \\
  $e_0$ & 1.101       &   2.242     \\
  $e_1$ & 7.922       & 14.798
\enddata
\tablecomments{$d_i$'s and $e_i$'s are defined through equation 
(\ref{eq_crit_nu}).}
\label{tab_nu_app}
\end{deluxetable}

\begin{deluxetable} {rrrrcrrrcrrr}  \tabletypesize{\scriptsize}
\tablewidth{0pt}  
\tablecaption{
Fitting coefficients for $\Delta\log\keff$ 
}
\tablehead { 
& \multicolumn{3}{c}{n=1.5} &
& \multicolumn{3}{c}{n=3}&
& \multicolumn{3}{c}{n=3.5}\\
\cline{2-4}\cline{6-8}\cline{10-12}\\
$\mu$&     $a_1$ &   $a_2$ &  $a_3$&
     &     $a_1$ &   $a_2$ &  $a_3$&
     &     $a_1$ &   $a_2$ &  $a_3$ }
\startdata
                                                                              
0.1  &0.73385   & 34.10425   &$-148.77090$  &
     &-0.03012  & 24.47968   &$-196.59523$  &
     &-0.08969  & 21.17797   &$-182.98519$  \\
0.2  & 0.88514  & 30.49465   &$-120.51368$  &
     & 0.12028  & 21.15117   &$-167.29413$  &
     & 0.07023  & 16.73734   &$-137.84158$  \\
0.3  & 1.00745  & 25.75547   &$-87.11816 $  &
     & 0.23724  & 17.77352   &$-135.98160$  &
     & 0.17016  & 14.06672   &$-113.16240$  \\
0.4  & 1.05899  & 23.41619   &$-91.19969 $  &
     & 0.32678  & 14.69585   &$-107.93152$  &
     & 0.24079  & 11.91660   &$-93.63182 $  \\
0.5  & 1.16935  & 15.72668   &$-29.29898 $  &
     & 0.41656  & 10.91861   &$-74.44114 $  &
     & 0.29680  & 10.04001   &$-79.29394 $  \\
0.6  & 1.19753  & 11.16234   &$-9.88845  $  &
     & 0.48402  &  7.43452   &$-45.76199 $  &
     & 0.39296  &  5.57693   &$-38.07912 $  \\
0.7  & 1.15145  &  8.27121   &$-8.49183  $  &
     & 0.52890  &  4.11300   &$-21.09957 $  &
     & 0.42220  &  3.49667   &$-23.59086 $  \\
0.8  & 1.04939  &  4.77987   &$ 1.27992  $  &
     & 0.52238  &  1.64450   &$-5.99889  $  &
     & 0.44813  &  0.61004   &$-3.06641  $  \\
0.9  & 0.79149  &  2.31621   &$ 4.14140  $  &
     & 0.40330  &  0.52695   &$-1.00724  $  &
     & 0.34866  &$-0.12201$  &$ 0.77103  $  \\
1.0  & 0.10778  &  1.67357   &$-2.14200  $  &
     & 0.04236  &  0.43492   &$-0.71212  $  &
     & 0.03456  &  0.28929   &$-0.54041  $

\enddata  
\tablecomments{$a_i$'s are defined through equation (\ref{eq_kcor_appr}).}
\label{tab_k2eff}
\end{deluxetable} 

\clearpage

\clearpage

\clearpage

\clearpage

\clearpage

\begin{thebibliography}{}

\bibitem[Anand(1968)]{ana68}      
  Anand, S.\ P.\ S.\ 1968, \apj, 153, 135
\bibitem[Arbutina(2009)]{arb09}
  Arbutina, B.\ 2009, \mnras, 394, 501
\bibitem[Baumgarte \& Shapiro(1999)]{bau99}
  Baumgarte, T.\ W., \& Shapiro, S.\ L.\ 1999, \apj, 526, 937
\bibitem[Binney \& Tremaine(2008)]{bin08}
  Binney, J., \& Tremaine, S.\ 2008, Galactic Dynamics, 2nd ed.\ 
  (Princeton: Princeton Univ.\ Press), 81
\bibitem[Bozkurt \& De\u{g}irmenci(2007)]{boz07}
  Bozkurt, Z., \& De\u{g}irmenci, \"{O}.\ L.\ 2007, \mnras, 379, 370
\bibitem[Brooker \& Olle(1955)]{bro55}
  Brooker, R.\ A., \& Olle, T.\ W.\ 1955, \mnras, 115, 101
\bibitem[Burrows \& Liebert(1993)]{bur93}
  Burrows, A., \& Liebert, J.\ 1993, Rev.\ of Mod.\ Phy., 65, 301
\bibitem[Carciofi et al.(2008)]{car08}
  Carciofi, A.\ C., Domiciano de Souze, A., Magalh\~aes, A.\ M., 
  Bjorkman, J.\ E., \& Vakili, F.\ 2008, \apj, 676, L41
\bibitem[Chandrasekhar(1933a)]{cha33a}   
  Chandrasekhar, S.\ 1933, \mnras, 93, 390
\bibitem[Chandrasekhar(1933b)]{cha33b}   
  Chandrasekhar, S.\ 1933, \mnras, 93, 449
\bibitem[Chandrasekhar(1933c)]{cha33c}   
  Chandrasekhar, S.\ 1933, \mnras, 93, 462
\bibitem[Chandrasekhar(1939)]{cha39}
  Chandrasekhar, S.\ 1939, An introduction to the study of stellar structure 
  (Chicago Univ. Press)
\bibitem[Claret(1999)]{cla99}     
   Claret, A.\  1999, \aap, 350, 56
\bibitem[Claret \& Gim\'enez (1993)]{cla93}     
   Claret, A., \& Gim\'enez, A.\ 1993, \aap, 227, 487
\bibitem[Claret \& Willems(2003)]{cla03}     
   Claret, A., \& Willems, B.\ 2003, \aap, 410, 289
\bibitem[Clement(1974)]{cle74}
   Clement, M.\ J.\ 1974, \apj, 194, 709
\bibitem[Cowling(1938)]{cow38}    
   Cowling, T.\ G.\ 1938, \mnras, 98, 734
\bibitem[Collins(1963)]{col63}
  Collins, G.\ W.\ 1963, \apj, 128, 1134
\bibitem[Domiciano de Souza et al.(2003)]{dom03}
  Domiciano de Souza, A., Kervella, P., Jankov, S., Abe, L.,
  Vakili, F., di Folco, E., \& Paresce, F.\ 2003, \aap, 407, L47
\bibitem[D'Souza et al.(2006)]{dso06}
  D'Souza, M.\ C.\ R., Motl, P.\ M., Tohline, E., \& Frank, J.\
  2006, \apj, 643, 381
\bibitem[Durney \& Roxburgh(1970)]{dur70}     
  Durney, B.\ R., \& Roxburgh, I.\ W.\ 1970, \mnras, 148, 239
\bibitem [Eggleton(1983)]{egg83}   
   Eggleton, P.\ P.\ 1983, \apj, 268, 368
\bibitem [Eggleton(2006)]{egg06}
   Eggleton, P.\ P.\ 2006, Evolutionary Processes in Binary and Multiple Stars
   (Cambridge: Cambridge Univ.\ Press)
\bibitem[Frank et al.(2002)]{fra02}
   Frank, J., King, A. R., \& Raine, D.\ J.\ 2002, Accretion Power in 
   Astrophysics, 3rd ed.\ (Cambridge: Cambridge Univ.\ Press)
\bibitem[Geroyannis et al.(1979)]{ger79} 
   Geroyannis, V.\ S., Tokis, J.\ N., \& Valvi, F.\ N.\ 1979, \apss, 64, 359
\bibitem[Geroyannis \& Valvi(1984)]{ger85} 
   Geroyannis, V.\ S., \& Valvi, F.\ N.\ 1985, \apj, 299, 695
\bibitem[Geroyannis(1988)]{ger88} 
   Geroyannis, V.\ S.\ 1988, \apj, 327, 273
\bibitem[Gim\'enez(1985)]{gim85}
   Gim\'enez, A.\ 1985, \apj, 297, 405
\bibitem[Gim\'enez(2007)]{gim07}
   Gim\'enez, A.\ 2007, in IAU Symp.\ 240: 
   Binary Stars as Critical Tools \& Tests in Contemporary Astrophysics, 
   eds. W.I.\ Hartkopf, E.F.\ Guinan, \& P. Harmanec (Cambridge: Cambridge 
   Univ.\ Press), 290
\bibitem[Gneden(2003)]{gne03}
   Gneden, O.\ Y.\ 2003, \apj, 589, 752
\bibitem [Hachisu(1986a)]{hac86a}    
  Hachisu, I.\  1986a, \apjs, 61, 479
\bibitem[Hachisu(1986b)]{hac86b}    
  Hachisu, I.\  1986b, \apjs, 62, 461
\bibitem[Hachisu et al.(1986a)]{hacetal86a}    
  Hachisu, I., Eriguchi, Y., \& Nomoto, K.\ 1986a, \apj, 308, 161
\bibitem[Hachisu et al.(1986b)]{hacetal86b}    
  Hachisu, I., Eriguchi, Y., \& Nomoto, K.\ 1986b, \apj, 311, 214
\bibitem[Hernquist \& Ostriker(1992)]{hern92}
  Hernquist, L., \& Ostriker, J.\ P.\ 1992, \apj, 386, 375
\bibitem[Hernquist et al.(1995)]{her95}
  Hernquist, L., Sigurdsson, S., \& Bryan, G.\ L.\ 1995, \apj, 446, 717
\bibitem[Hjellming \& Webbink(1987)]{hje87}
  Hjellming, M.\ S., \& Webbink, R.\ F.\ 1987, \apj, 318, 794
\bibitem [Horedt(2004)]{hor04}
  Horedt, G.\ P.\ 2004, Polytropes : Applications in Astrophysics and 
  Related Fields, Astrophysics and Space Science Library Vol.\ 306 
  (Kluwer Academic Publishers: Dordrecht), 198
\bibitem [Hurley \& Roberts(1964)]{hur64}
  Hurley, M., \& Roberts, P.\ H.\ 1964, \apj, 180, 583
\bibitem[Jackson(1970)]{jac70}    
   Jackson, S.\ 1970, \apj, 160, 685
\bibitem[Jackson et al.(2004)]{jac04}    
   Jackson, S., MacGregor, K.\ B., \& Skumanich, A.\ 2004, \apj,  606, 1196
\bibitem[Jackson et al.(2005)]{jac05}    
   Jackson, S., MacGregor, K.\ B., \& Skumanich, A.\ 2005, \apjs,  156, 245
\bibitem[James(1962)]{jam62}
   James,  I.\ 1962, Ph.D. Thesis, Manchester University
\bibitem[James(1964)]{jam64}
   James,  R.\ A.\ 1964, \apj,  140, 552
\bibitem[Khaliullin \& Khaliullina(2007)]{kha07}
   Khaliullin, Kh.\ F., \& Khaliullina, A.\ I.\ 2007, \mnras, 382, 356
\bibitem[Kopal(1972)]{kop72}  
   Kopal, Z.\ 1972, Adv.\ Astron.\ Ap.\ 9, 1
\bibitem[Kopal(1978)]{kop78}  
   Kopal, Z.\ 1978, Dynamics of Close Binary Systems (Reidel: Dordrecht)
\bibitem[Kopal(1989)]{kop89} 
   Kopal, Z.\ 1989, The Roche Problem (Kluwer Academic Publishers: Netherlands)
\bibitem[Kippenhahn \& Thomas (1970)]{kip70} 
   Kippenhahn, R., \& Thomas, H-C.\ 1970, Stellar Rotation, 
   ed. A.\ Slettebak, I.A.U.\ Coll.\ 4, (Reidel: Dordrecht)
\bibitem[Lal et al.(2006)]{lal06}        
   Lal, A.\ K., Saini, S., Mohan, C., \& Singh, V.\ P.\ 2006, \apss, 306, 165
\bibitem[Linnell(1977)]{lin77}    
   Linnell, A.\ P.\ 1977,  \apss, 48,  165
\bibitem [Linnell(1981)]{lin81}    
   Linnell, A.\ P.\ 1981,  \apss, 80,  501
\bibitem[MacGregor et al.(2007)]{mac07}
   MacGregor, K.\ B., Jackson, S., Skumanich, A., \& Metcalfe, T.\ S.\ 
   2007, \apj, 663, 560
\bibitem[Mark(1968)]{mar68}
   Mark, J.\ W-K.\ 1968, \apj, 154, 627
\bibitem[Malkov(1993)]{mal93}     
   Malkov, O.\ Yu.\ 1993, Bull.\ Inf.\ CDS, 42, 27
\bibitem[Malkov(2003)]{mal03}     
   Malkov, O.\ Yu.\ 2003,  \aap, 402, 1055
\bibitem[Malkov(2007)]{mal07}     
   Malkov, O.\ Yu.\ 2007,  \mnras, 382, 1073
\bibitem[Martin(1970)]{mar70}     
   Martin, P.\ G.\ 1970,  \apss, 7, 119
\bibitem[McNally(1965)]{mcn65}
   McNally, D.\ 1965, The Observatory, 85, 166
\bibitem[Meynet \& Maeder(1997)]{may97}     
   Meynet, G., \& Maeder, A.\ 1997, \aap, 321, 465
\bibitem[Mohan et al.(1992)]{moh92}
   Mohan, C., Lal, A.\ K., Singh, V.\ P.\ 1992, \apss, 193, 69
\bibitem[Mohan et al.(1997)]{moh97}
   Mohan, C., Lal, A.\ K., \& Singh, V.\ P.\ 1997, \apss, 254, 97
\bibitem[Mohan \& Saxena(1983)]{moh83}
   Mohan, C., \& Saxena, R.\ M.\ 1983, \apss, 95, 369
\bibitem[Mohan et al.(1990)]{moh90}
   Mohan, C., Saxena, R.\ M., \& Agarwal, S.\ R., 1990,\apss 163, 23 
\bibitem[Mohan \& Singh(1978)]{moh78}
   Mohan, C., \& Singh, V.\ P.\ 1978, \apss 54, 293
\bibitem[Monaghan \& Roxburgh(1965)]{rox65}   
   Monaghan, F.\ F., \& Roxburgh, I.\ W.\ 1965, \mnras, 131, 13
\bibitem[Motl et al.(2007)]{mot07}
   Motl, P.\ M., Frank, J., Tohline, J.\ E., \& D'Souza, M.\ C.\ R.\
   2007, \apj, 670, 1314
\bibitem[Murray \& Dermott(1999)]{mur99}
  Murray, C.\ D., \& Dermott, S.\ F.\ 1999, Solar System Dynamics
  (Cambridge: Cambridge Univ.\ Press) 264
\bibitem[Naylor \& Anand (1972a)]{nay72a}
   Naylor, M.\ D.\ T, \& Anand, S.\ P.\ S.\ 1972a,  \apss, 16, 137 
\bibitem[Naylor \& Anand (1972b)]{nay72b}
   Naylor, M.\ D.\ T, \& Anand, S.\ P.\ S.\ 1972b,  \apss, 18, 59 
\bibitem[Ostriker \& Mark(1968)]{ost68}   
   Ostriker, J.\ P., \& Mark, J.\ W-K.\ 1968, \apj, 151, 1075
\bibitem[Ostriker \& Bodenheimer(1968)]{ost_bod68}   
   Ostriker, J.\ P., \& Bodenheimer, P.\ 1968, \apj, 151, 1089
\bibitem[Paczy\'nski(1971)]{pac71}
   Paczy\'nski, B.\ 1971, \araa, 9, 183
\bibitem[Paczy\'nski(1976)]{pac76}
   Paczy\'nski, B.\ 1976, IAU Symposium 73, Structure and Evolution of
   Close Binary Systems, ed.\ P.\ Eggleton, S.\ Mitton, \& J.\ Whelan
   (Reidel: Dordrecht), 75
\bibitem[Pasinetti-Fracassini et al.(2001)]{pas01}  
   Pasinetti-Fracassini, L.\ E., Pastori, L., Covino, S., \& Pozzi, A.\ 
   2001, \aap, 367, 521
\bibitem[Press et al.\ (1988)]{pre88}      
   Press, W.\ H., Teukolsky, S.\ A., Vetterling, W.\ T., \& 
   Flannery, B.\ P.\ 1988, Numerical Recipes in C, 2nd ed.\ 
   (Cambridge: Cambridge Univ.\ Press)
\bibitem[Quataert et al.(1996)]{qua96}
   Quataert, E.\ J., Kumar, P., \& Ao, C.\ O.\ 1996, \apj, 463, 284
\bibitem[Rappaport et al.(1982)]{rap82} 
   Rappaport, S., Joss, P.\ C., \& Webbink, R.\ F.\ 1982, \apj, 254, 616
\bibitem[Ribas(2006)]{rib06}      
   Ribas, I.\ 2006, \aaps, 304, 89
\bibitem[Russell(1928)]{rus28}    
   Russell, H.\ N.\ 1928, \mnras, 88, 641
\bibitem[Schwarzschild(1958)]{sch58}
   Schwarzschild, M.\ 1958, Structure and Evolution of the Stars
   (Princeton: Princeton Univ.\ Press)
\bibitem[Singh \& Singh(1983)]{sin83}
   Singh, M., \& Singh, G.\, 1983, \apss, 96, 313
\bibitem[Singh \& Singh(1984)]{sin84}     
   Singh, M., \& Singh, G.\ 1984, \apss, 106, 161
\bibitem[Stahler(1983a)]{sta83a}
   Stahler, S.\  W.\ 1983a, \apj, 268, 155
\bibitem[Stahler(1983b)]{sta83b}
   Stahler, S.\  W.\ 1983b, \apj, 268, 165
\bibitem[Stern(1939)]{ste39}      
  Stern, T.\ E., 1939, \mnras, 99, 451
\bibitem[Stothers(1974)]{sto74}     
  Stothers, R.\ 1974, \apj, 194, 651
\bibitem[Smeyers \& Willems(2001)]{sme01}
  Smeyers, P., \& Willems, B.\ 2001, \aap, 373, 173
\bibitem[Tassoul(1987)]{tas87}
  Tassoul, J.\ L.\ 1987, \apj, 332, 856
\bibitem[Tassoul(1988)]{tas88}
  Tassoul, J.\ L.\ 1988, \apj, 324, L71
\bibitem[Torres et al.(2006)]{tor06}     
  Torres, G., Lacy, C.\ H., Marschall, L.\ A., Sheets, H.\ A., \& 
  Mader, J.\ A.\ 2006, \apj, 640, 1018
\bibitem[Townsend et al.(2004)]{tow04}
  Townsend, R.\ H.\ D., Owocki, S.\ P., \& Howarth, I.\ D.\ 2005, 
  \mnras, 350, 189
\bibitem[Weinberg(1999)]{wei99}
  Weinberg, M.\ D.\  1999, \aj, 117, 629
\bibitem[Wolff et al.(2006)]{wol06}
  Wolff, S.\ C., Strom, S.\ E., Dror, D., Lanz, L., \& Venn, K.\ 2006,
  \aj, 132, 749
\bibitem[Zahn(1977)]{zah77}
  Zahn, J.\ P.\ 1977, \aap, 57, 383
\end{thebibliography}
\end{document}